\documentclass[]{aa}
\pdfoutput=1
\usepackage{color}
\usepackage{natbib}
\usepackage[autostyle]{csquotes}
\bibpunct{(}{)}{;}{a}{}{,} 

\def\msun{$\mbox{M}_\odot$}

\def\afe{{[$\alpha$/Fe]}}
\def\feh{{[Fe/H]}}
\def\ageo{{$\mathrm{age}_\mathrm{o}$}}
\def\afeo{{[$\alpha$/Fe]$_\mathrm{o}$}}
\def\feho{{[Fe/H]$_\mathrm{o}$}}
\def\Zo{{Z$_\mathrm{o}$}}
\def\agei{{$\mathrm{age}_\mathrm{i}$}}
\def\afei{{[$\alpha$/Fe]$_\mathrm{i}$}}
\def\fehi{{[Fe/H]$_\mathrm{i}$}}
\def\Zi{{Z$_\mathrm{i}$}}

\begin{document}

\title{Abundance patterns in early-type galaxies: is there a 'knee' in the \feh~vs.~\afe~relation?}
\subtitle{}

\author{C. J.~Walcher\inst{\ref{i:CJW}}; P.R.T.~Coelho\inst{\ref{i:PC}}; A.~Gallazzi\inst{\ref{i:AG},\ref{i:AG2}}; G.~Bruzual\inst{\ref{i:GB}}; S.~Charlot\inst{\ref{i:SC}}; C. Chiappini\inst{\ref{i:CJW}}}

\institute{
Leibniz-Institut f\"ur Astrophysik Potsdam (AIP), An der Sternwarte 16, 14482 Potsdam, Germany \label{i:CJW}
\and
Instituto de Astronomia, Geof\'isica e Ci\~encias Atmosf\'ericas, Universidade de S\~ao Paulo, Rua do Mat\~ao 1226, 05508-090 - S\~ao Paulo - Brasil \label{i:PC}
\and
INAF-Osservatorio Astrofisico di Arcetri, Largo Enrico Fermi 5, I-50125 Firenze, Italy \label{i:AG}
\and 
Dark Cosmology Center, University of Copenhagen, Niels Bohr Institute, Juliane Maries Vej 30, 2100 Copenhagen, Denmark \label{i:AG2}
\and
Centro de Radioastronom\'ia y Astrof\'isica (CRyA), Morelia, Michoacan 58089, Mexico \label{i:GB}
\and
UPMC-CNRS, UMR7095, Institut d'Astrophysique de Paris, F-75014 Paris, France \label{i:SC}
}

\authorrunning{Walcher, et al.}
\titlerunning{Enrichment histories of ETGs}
\date{Received date / Accepted date }

\abstract{Early-type galaxies (ETGs) are known to be enhanced in $\alpha$ elements, in accordance with their old ages and 
short formation timescales.
In this contribution we aim to resolve the enrichment histories of ETGs. This means we study the abundance of Fe 
(\feh) and the $\alpha$-element groups (\afe) separately for stars older than 9.5 Gyr (\feho, \afeo) and for stars between 1.5 
and 9.5 Gyr (\fehi, \afei).
Through extensive simulation we show that we can indeed recover the enrichment history \emph{per galaxy}. We then 
analyze a spectroscopic sample of 2286 early-type galaxies from the SDSS selected to be ETGs. We separate out those 
galaxies for which the abundance of iron in stars grows throughout the lifetime of the galaxy, i.e.~in which \feho~$<$ \fehi. 
We call those consistent with self-enrichment, while the others must have experienced  
some mergers or significant gas accretion.
We confirm earlier work where the \feh~and \afe~parameters are correlated with the mass and velocity dispersion of ETGs. 
We emphasize that the strongest relation is between \afe~and age. This relation falls into two regimes, one with a steep 
slope for old galaxies and one with a shallow slope for younger ETGs. The vast majority of ETGs in our sample do not show 
the \enquote*{knee} in the plot of \feh~vs.~\afe~commonly observed in local group galaxies. This implies that for the vast majority of 
ETGs, the stars younger than 9.5 Gyrs are likely to have been accreted or formed from accreted gas. 
The properties of the intermediate-age stars in accretion-dominated ETGs indicate that mass growth through late 
(minor) mergers in ETGs is dominated by galaxies with low \feh~and low \afe. 
The method of reconstructing the stellar enrichment histories of ETGs introduced in this paper promises to constrain 
the star formation and mass assembly histories of large samples of galaxies in a unique way.}


\maketitle

\section{Introduction}
\label{s:intro}

The stars in massive early-type galaxies (ETGs) have had to form early in the history of the universe for two reasons: 1) Very 
few of these galaxies have (significant) young stellar populations \citep[e.g.,][]{yi05,trager09, thomas10} and 2) the ratio of 
$\alpha$ elements (O, Mg, etc.) to Fe in the atmospheres of these stars is higher than solar \citep[e.g.,][]{peterson76,worthey92, 
milone00}, indicating a short timescale for the 
enrichment of the ISM out of which they formed. The ratio \afe~is a powerful estimator of the duration of 
star formation \citep{tinsley79, matteucci86}. Study of the mass dependence of the ages and enhancement ratios of ETGs 
has led to the classical picture in which the star formation timescales are shorter with increasing galaxy mass 
\citep{thomas05}. 

Direct studies of massive galaxies at high redshift support this picture in which at least some ETGs must 
have "formed the bulk of their mass only a few million years after the big bang" \citep{nayyeri14}. On the other hand there 
is evidence that massive ETGs cannot be described by a stellar population with a single age and abundance pattern 
\citep[recently, e.g.,][]{lonoce14}. Indeed, \citet{arimoto86} emphasize that this is true in terms not only of an age spread, 
but also of an abundance spread: "The present analysis shows that such metal-poor stars in the RGB evolution have 
a strong influence on the integrated colors of the galaxy."

Recent observational and theoretical work has focused on the question of when and how the late mass growth of early-type 
galaxies takes place. Two opposite possibilities for such mass growth exist: 1) in situ star formation from gas that was accreted in the initial 
phase of the galaxies formation and 2) mass growth from later major or minor dry merging. A third possibility with less clear 
observational signatures is 3) stellar mass growth through star formation from gas that was either accreted during a merger 
(wet mergers) or in a cold mode from the cosmic web. 

Different observational approaches to these questions exist. An important piece of the puzzle is the observation that the 
progenitors of ETGs at redshifts above z$\sim$1 are compact, dense objects \citep{trujillo06, van-dokkum08,van-der-wel08}. 
\cite{naab09}, \citet{hopkins09}, and \citet{van-dokkum10} show that the cores of today's ellipticals are consistent with being the dense 
galaxies observed at high redshift, although \citet{van-der-wel11} make the point that the majority of these distant, dense galaxies 
should be thought of as disk-like, introducing some tension with the mostly dispersion-dominated cores of today's early-type galaxies. 
Overall, the data have led to the notion that (minor) merging over a Hubble time will simultaneously 
contribute to the mass growth of these objects and to their size growth 
\citep{clemens09, naab09, van-dokkum10, oser12, greene12, greene13}. 

Simulations predict that stellar accretion is most important at large radii \citep{lackner12, navarro-gonzalez13, hirschmann14}. 
In contrast to earlier notions, simulations even show that no major merger at all is needed to 
form massive ellipticals \citep{bournaud07}. On the other hand, direct measurement of major merging is 
available in the literature \citep[e.g.,][]{bell06, de-ravel09, robaina10, robotham14}. These studies find that the typical 
L* galaxy will have undergone one major merger after redshift one \citep{keenan14} and that 
the higher the mass of a galaxy, the more it grows through major merging. The evolution in the mass function can be 
combined with the expected growth due to star formation to yield a direct measure of the mass growth due to mergers in 
ETGs \citep[][]{walcher08, ownsworth14}. This leads to estimates of roughly half of the present-day total stellar mass of ETGs being 
formed in other galaxies prior to their incorporation into the ETG over cosmic time. However, mergers also contribute gas. 
For example \cite{kormendy09} discuss how structural 
properties can be used to distinguish between dry mergers that would contribute only stars and wet mergers that would 
in turn contribute gas, thus leading to late star formation and rebuilding of cusps. 

The present paper is dedicated to measuring the relative importance of the mass-growth processes from their chemical signature. 
In particular, other observational approaches cannot hope to distinguish between in situ formation of stars from 
gas left over from any putative "initial accretion", on the one hand, and gas subsequently accreted either in a cold mode or through minor 
mergers, on the other. This distinction is much easier to study from chemical signatures as shown below. 

\begin{figure}[tbp]
\begin{center}
\resizebox{1\hsize}{!}{\includegraphics[]{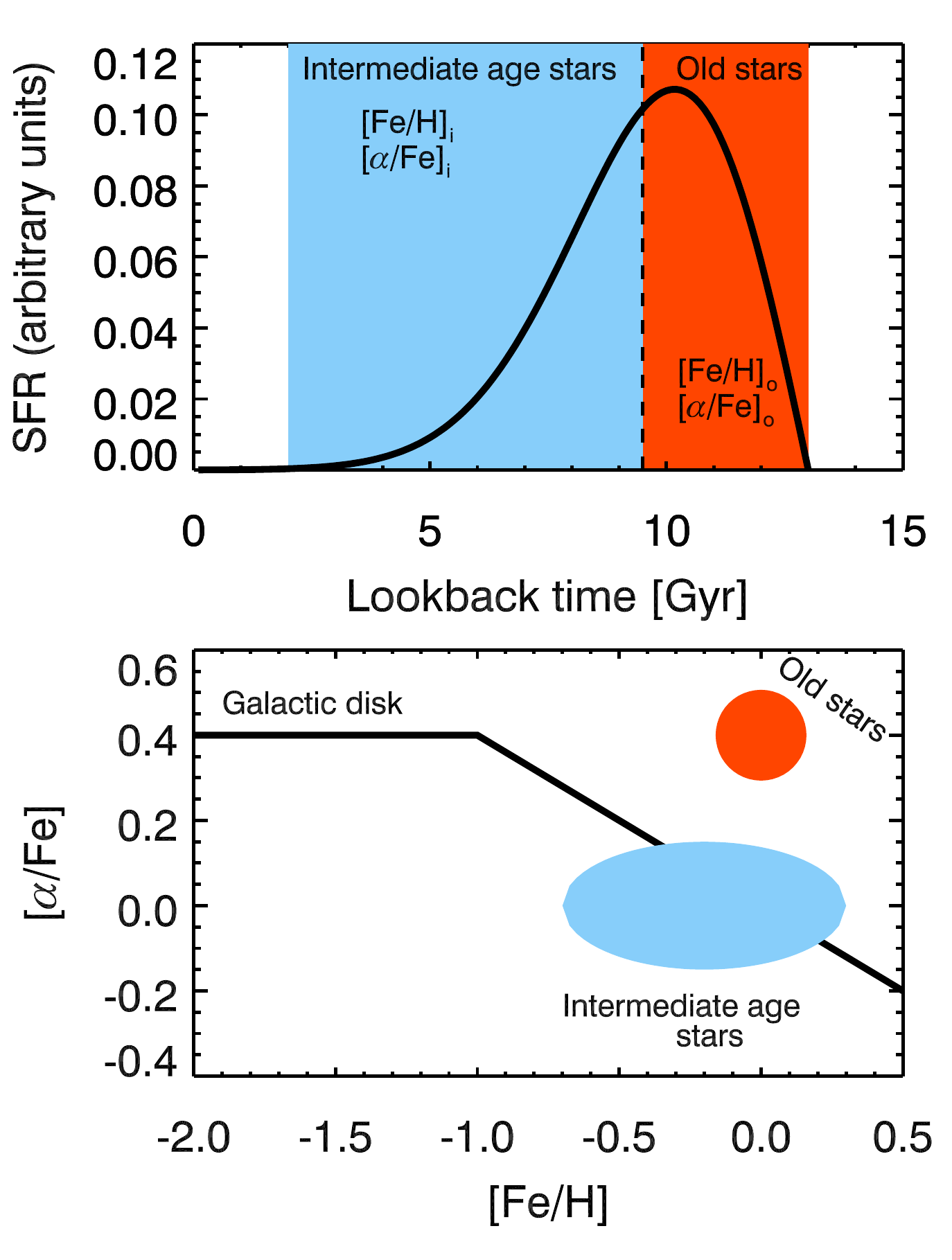}}
\end{center}
\caption[didactic]{Schematic representation of the new parameters introduced in this paper, namely \afeo~and \feho. These represent 
the average \afe~and \feh~of stars older than 9.5 Gyr on a per galaxy basis. The upper panel shows a schematic star formation history, 
which separates into these two stellar populations. The lower panels depicts on the standard \afe~vs. \feh~diagram familiar for 
the Milky Way what is expected for the old and intermediate age stars respectively. }
\label{f:didactic}
\end{figure}

In this contribution we define the "stellar enrichment histories" (SEH) as the four-dimensional space occupied by the stellar populations in each 
galaxy and spanned by the parameters age (i.e., time elapsed since formation of those populations), \feh, \afe~and 
the mass (or luminosity) contribution of these stars. It needs to be kept in mind that we can only address the stellar population 
within one spatial resolution element (one fiber in our case, the case of SDSS). Because of measurement uncertainties 
we very strongly bin this space into just two age bins: the old age bin contains all stars older than 9.5Gyr, the intermediate 
age bin contains all stars between 1.5 and 9.5 Gyr of age. Within each age bin and for each galaxy we can compute the 
luminosity-averaged \feh~and \afe~as well as the fractional contribution of the old and intermediate populations to the total 
luminosity of each galaxy. This principle 
is illustrated in the upper panel of Figure \ref{f:didactic}. For the remainder of this paper we denote the mean \afe~and \feh~of the 
old stars with the subscript 'o' (i.e.~we write \feho~and \afeo), whereas we denote the mean \afe~and \feh~of the intermediate 
age stars with the subscript 'i' (i.e.~we write \fehi~and \afei). The "star formation history" (SFH) would simply be the marginalization 
of the SEH onto the 2-dimensional space of age and luminosity contribution. 

The choice of 9.5 Gyr for separating old and intermediate stellar populations in ETGs is of course to some level arbitrary. 
It can, however, be justified on several grounds as providing the widest age bin for old stars that still 
makes sense physically. On the one hand simulations such as those of \citet{naab09} show the 
typical mass assembly histories of ETGs in a hierarchical universe. For many of these galaxies there is a change of regime at 
around 10 Gyr ago, where the mass build-up changes from being dominated by in-situ star formation to accretion of other, 
smaller galaxies and clumps. Thus this is an age where we expect to maximize the signal from any difference between old 
and intermediate age population. On the other hand, those alpha-enhanced populations that we can study in detail 
in the Milky Way are all older than 11 Gyr. This is true for the thick disk \citep{fuhrmann11} and for the bulge 
\citep{bensby13}. As shown later in the present paper (Figure 12, lower right panel) and in a paper in preparation 
(Walcher et al., in prep.), this effect is seen in the luminosity-weighted average properties at a lookback time of around 9.5 Gyr 
as well in the sense that the rate of change of \afe~with time shows a significant change of slope. These 2 or 3 Gyr 
after the onset of star formation are thus different in the chemical signatures, which helps for the technical as 
well as scientific arguments of the present paper. 
Finally, the cosmic star formation density peaks at a lookback time of 10 Gyr, as shown in the recent review by 
\citet[][their Figure 9]{madau14}. Specifically for ETGs, this is also the time when quiescent 
galaxies started appearing in large number \citep[see e.g.,][, their Figure 13]{arnouts07}. 

The lower panel of Figure \ref{f:didactic} illustrates basic expectations for the locus the old and intermediate stellar population 
would occupy in the projection onto the \afe~vs.~\feh~plane. As is well known the 'knee' in this diagram carries information about the 
star formation efficiency in the galaxy, where higher star formation efficiency leads to higher overall metallicity before the onset 
of SNIa enrichment, and thus to a knee that is located at higher \feh~values. The low-\feh~plateau in \afe~can be modulated by 
the slope of the upper end of the stellar initial mass function, but the data we discuss here do not allow any inference concerning 
this, thus we neglect this question for the remainder of the paper. 

The existence of the knee in the lower panel of Figure \ref{f:didactic} for the Milky Way stars is tied to some causal connection at each 
time step during the star formation history between the existing stars and the next generation of stars. Although some accretion is 
needed, if only to solve the g-dwarf question, we generally call this a scenario that is consistent with self-enrichment (SLF). 
The SLF scenario would predict in our specific case of coarsely resolved enrichment histories for ETGs that the \feh~of the 
intermediate-age population would always be higher than for the old population and the \afe~of the intermediate-age 
population would always be lower than that of the old population. On the other hand, as mentioned above we know that mergers 
contribute a significant percentage of the stars in massive ETGs. The merging stars bear no connection to the pre-existing stars in 
the ETG and thus could have any \feh~and \afe~properties. The shorthand for this scenario throughout this paper is 
ACC for accretion. 

The mass-metallicity relation 
\citep{tremonti04, gallazzi05} leads to the expectation that minor, i.e.~ low mass mergers would contribute stars with sub-solar 
\feh. The \afe~of these stars is less certain. However, \cite{sansom08} show that the high \afe~abundances seen in massive 
quiescent galaxies do not continue to the lower luminosity quiescent galaxies, as those have generally lower \afe. Also, lower 
mass galaxies tend to have prolonged star formation histories, again leading to an expectation of lower \afe~in general. We 
caution that we currently do not know what the \afe~of intermediate-redshift minor mergers could be. Nevertheless, we 
are led to an expectation that the intermediate-age stars in ETGs would on average have solar \afe~and \feh~either slightly 
above or even significantly below the solar value (the blue ellipse in Figure \ref{f:didactic}). Indeed, \cite{coelho09} found that for the galaxy M32 
the metal-poor component of their stellar population fit corresponds to a younger population than the metal-rich one. 

Although simple, this picture can be used to provide a framework for interpreting the results in this paper. Complications arise, 
however, that need to be acknowledged. For example, \cite{pipino09} show that semi-analytic models cannot reproduce 
the \afe-mass relations yet, owing to inadequate treatment of late mergers, thus clearly implying that we do not have sufficient 
theoretical understanding of low-mass galaxy evolution in simulations. More generally, \citet{hirschmann13} conclud that, 
while it is clear that accretion, minor merging, major merging, star formation efficiency, and winds must all work together to form 
those galaxies we see today, their exact interplay is still not sufficiently understood. Simulations thus do not produce a 
satisfying population of galaxies overall. Observationally, and in the direct context of this contribution, this is 
illustrated by the recent work of \cite{renzini14}, who concluded from data on the enrichment of the intracluster medium 
"that even the most massive galaxies must have lost a major fraction of the metals they have produced". 

An uncertainty that much more directly affects our distinction in SLF and ACC ETGs has been spelled out in the works of 
\citet{arimoto86} and \citet{vazdekis96}. Indeed, it turns out that metallicity may not monotonously increase with age for the most 
massive ETGs even in absence of accretion events. 
After a peak in metallicity at a certain age, this metallicity starts declining again because of the late mass loss by the 
abundant low-metallicity stars still present in the galaxy. Comparison with Figure 11 of \citet{vazdekis96} reveals that the 
peak in metallicity may be reached very rapidly (within 2 or 3 Gyrs after formation of the galaxy). However, the decline in 
metallicity remains subtle and is of order 0.2 dex in [Z] until redshift zero. We come back to this value in Section \ref{s:dichotomy}.

The present contribution introduces extensions to the method of spectral fitting. We therefore present our algorithm in detail in 
Section \ref{s:models} and then spend significant space on understanding any possible method-intrinsic effects, such as degeneracies, 
in Section \ref{s:simul}. Finally we also verify in detail that the method is applicable to the data at hand in Section \ref{s:fit1}. 
Readers wishing to read about the scientific results first can jump directly to Section \ref{s:results}. 

\section{Sample}
\label{s:sample}

For this paper we are interested in a significant sample of early-type galaxies (ETGs) with high S/N spectra. With ETG we mean 
a "spectral ETG", i.e.~essentially a galaxy with a spectrum dominated by old stars. Such a sample is easily selected from 
the MPA/JHU catalogs\footnote{\url{http://home.strw.leidenuniv.nl/~jarle/SDSS/}} for SDSS DR7  \citep{abazajian09}, containing general 
information about the galaxy's spectrum (S/N, stellar velocity dispersion), emission lines and absorption indices, as well as photometric 
information. ETGs in the SDSS have been studied in many contributions to the literature \citep[e.g.,][]{bernardi03,zhu10}. The 
specific reason to revisit these spectra is that they provide a large sample of readily available high quality spectra with which 
to test the power of the new models and the new methodology we apply here. In particular we adopt the following criteria to 
select our sample:\\

\begin{itemize}
\item Signal to noise larger than 40
\item r-band concentration index larger or equal to 2.8
\item "unclassifiable", i.e. with S/N \textless 3 in all of the BPT diagram emission lines, thus excluding AGN
\item Velocity dispersion between 40 and 375 km/s
\item Eliminate duplicate observations, keeping the one with higher S/N
\end{itemize}

We further estimate stellar masses from absorption indices following the exact same method as \citet{gallazzi05}. The sample is thus further 
restricted to those galaxies for which a spectroscopic stellar mass could be obtained (i.e. for which the set of five absorption features, 
$r-i$ fiber color and $z$-band model magnitude are available). We use the 16, 50, 84 percentiles of the probability distribution function 
for the lower 1$\sigma$ errorbar, the actual value and the upper 1$\sigma$ error bar, respectively. These masses are normalized to the 
$z$-band modelmag and a Chabrier IMF. 

\begin{figure}[tbp]
\begin{center}
\resizebox{1\hsize}{!}{\includegraphics[]{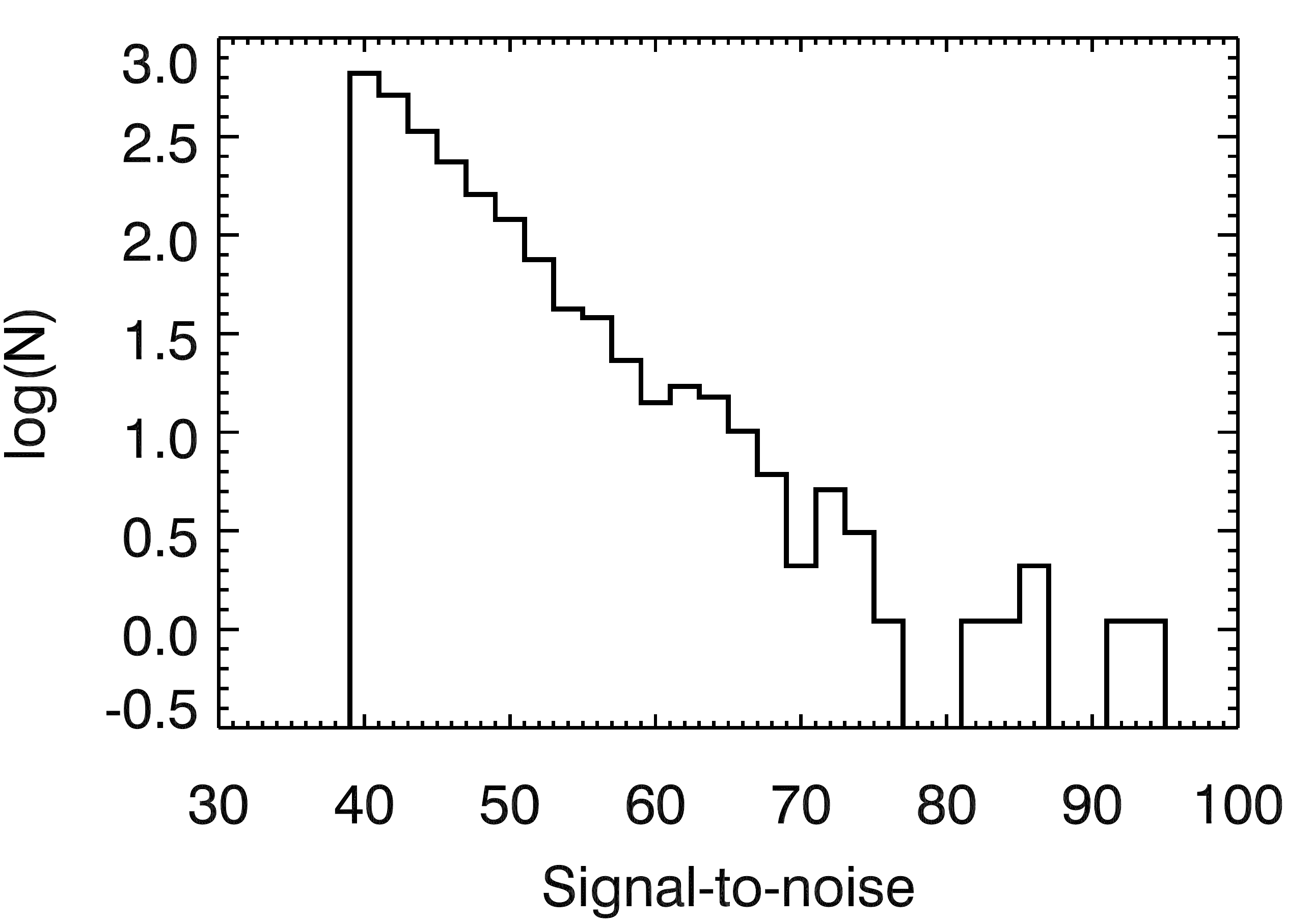}}
\end{center}
\caption[sn100]{S/N distribution of our sample galaxies. }
\label{f:ston}
\end{figure}

These criteria give us a sample of 2286 pure early-type galaxies. This sample has been S/N selected and it would thus be a daunting task 
to follow through all selection criteria to infer how representative the sample is for all early-type galaxies. We thus refrain from such an 
exercise. The S/N distribution of our sample is shown in Figure \ref{f:ston}. 

\section{Stellar population models and fitting procedure}
\label{s:models}

Stellar population models exploring the effects of varying element abundance ratios on specific line indices have been driving 
our understanding of ETG formation and evolution \citep[e.g.,][]{lee09}. In particular, studying the dependancies of single elements 
\cite[e.g.,][]{smith09, worthey14} is interesting in its own right, but is currently not possible for the goals of this paper for two reasons: 
1) We want to study SEHs in a resolved way and it would be difficult to control physical degeneracies. 2) We need spectral 
models with a restricted set of free parameters to keep our algorithms mathematically under control. Taking e.g.~the recent 
contribution by \cite{worthey14} as a comparison and referring back to Figure \ref{f:didactic} we can highlight one important  
point of our paper. These authors consider abundance distribution functions (ADFs) within each ETG to address the similar 
questions to the present contribution, namely how abundance ratios are spread within ETGs. However, due to the nature 
of the fitting algorithm \cite{worthey14} have to parameterize these ADFs, thus leading to a prior assumption on the SEHs 
possible in the galaxies they analyze. Because knowledge of the real SEHs of ETGs is so scarce we do not believe that 
putting such a prior is warranted yet. In support of our approach we also highlight that the use of \afe~as a single 
parameter involving a set of elements is also justified by the findings of \cite{johansson12}, who find that the correlations of 
[O/Fe], [Mg/Fe], [C/Fe] with velocity dispersion are identical for ETGs. In general, models capable of predicting the effects 
of $\alpha$-elements on the full optical spectrum of stellar populations are starting to become less rare, see \cite{coelho07}, 
\cite{koleva08c}, \cite{walcher09}, \cite{percival09}, \cite{vazdekis11}, \cite{conroy12}, \cite{prugniel12}, \cite{vazdekis15}. 

We use an updated version of the differential stellar population models presented in \citet[][hereafter W09]{walcher09} 
and we refer to that paper for a full description. In brief, to compute the models we combine the simple stellar population (SSP) 
spectra from fully theoretical stellar population models by \citet[][]{coelho07} \citep[based on isochrones from][]{weiss07} 
with semi-empirical models by \citet{vazdekis10} and \citet{falconbarroso11}. Of the \citet{vazdekis10}  models we only 
use those with solar metallicity 
and abundance ratios. The model spectra for non-solar \feh~and \afe~values are then obtained using
the C07 models of the same age as a prediction for the effects of varying \feh~and \afe. 
We show in W09 that these models yield reliable values for the ages, the \feh~and the \afe~
values of Milky Way globular clusters within the range covered by our models. 

During the course of the project we realized that the parameter coverage of the \cite{coelho07} models (age: 3-12 Gyr, 
\feh: $-$0.5, $-$0.25, 0.0, 0.2 and \afe: 0.0, 0.4) lead to "edge effects", in the sense that when fitting a stellar population 
older than 12 Gyr and with solar abundance ratios, age-metallicity degeneracy will drive the best-fitting model to have 
super-solar \feh~and \afe. We have therefore decided to extrapolate the grid of theoretical models to an age range of 
2-13 Gyr, \feh~values of $-$0.5, $-$0.25, 0.0, 0.2 
and \afe~values of $-$0.2, 0.0, 0.2, 0.4. The theoretical spectra are interpolated linearly in a multi-dimensional space with the 
axes: pixel, log(Flux), age, \feh~and \afe~before being applied to the semi-empirical models. 

Metallicity is defined as a combination of \feh~and \afe, explicitly for our models [Z] = \feh~+ 0.75 * \afe~\citep[see][]{coelho07}. 

Although we select galaxies for our analysis that have all signs of lacking young stellar populations, we cannot rule out any contribution. 
In the literature, rejuvenation of ETGs is still a controversial subject with findings related to sample definition, analysis methods and 
uncertainties related to late, hot phases of stellar evolution \citep[\feh~][]{yi05,sanchez-blazquez09, ocvirk10, thomas10}. 
We therefore extend our template set by five templates from \citet{vazdekis10} with younger ages (1, 0.63, 0.32, 0.1, 0.06 Gyr). 
These do not have varying \afe~ ratios. We include these templates into the fitting in all cases and use them to verify that 
indeed contributions by younger stellar populations are negligible in the data we analyze. 

Full spectrum fitting has survived a number of stringent tests \citep[\feh~][]{cid-fernandes05, koleva08a} and is 
becoming increasingly popular 
as a method to derive stellar population properties of galaxies \citep[\feh~][to cite but a few]{heavens00, cappellari04, cid-fernandes13}. 
For analysis of the spectra using full spectrum fitting we use the software {\tt paradise}. The underlying algorithm is fully 
described in Appendix \ref{s:paradise} and the same software was used in our previous papers \citep{walcher06, walcher09}.
In one sentence, {\tt paradise} fits model template spectra to observed spectra taking into account the uncertainties on each 
spectral pixel, solving for the kinematics of the stellar population and allowing very flexible masking of wavelength regions 
due to quality concerns. 


\section{Analysis of mock galaxy spectra}
\label{s:simul}

The goal of this paper is to shed new light on the SEHs of early-type galaxies to understand their formation history. Before we can 
do so, we must verify that our method is in principle capable of providing us with SEHs. In the present section we describe the creation 
of mock galaxy spectra and their analysis. As we use the same models for the mock spectra as we then use to analyze these, 
we can only test the accuracy and intrinsic degeneracies of our method under the assumption that the model is a perfect representation 
of the data. We performed an independent test of the accuracy 
of our models as compared to globular cluster spectra in \cite{walcher09}. In the remainder of the paper we report what we 
find when fitting the wavelength range 4828 to 5364A. This wavelength range has been shown in W09 to most accurately recover 
age, [Fe/H] and [alpha/Fe] for bulge globular clusters.

The main conclusion of this section is that we can recover the luminosity-weighted average properties of ETGs and the element abundance 
patterns of their old stars with good accuracy and precision. Recovering the properties of intermediate-age stars is a more 
challenging task.

\subsection{Mock galaxy spectra}
\label{s:simul_mock}

As motivated in the introduction, we analyze the SEHs of our galaxies in three bins of age $\tau$, namely $\tau \le$ 1.5 Gyr 
(called young age bin from now on), 1.5 Gyr $< \tau \le$ 9.5 Gyr (called intermediate-age bin, subscript i), $\tau >$ 9.5 Gyr 
(called old age bin, subscript o). We do not attempt a better time resolution for the reasons given below and because 
experience with full spectrum fitting shows that three bins in age are a reasonable assumption for the best age resolution 
one can hope for in the star formation history \citep[compare e.g.][]{cid-fernandes05,tojeiro07}. We therefore construct our mock galaxies 
by separating our SSPs in the three age bins above and assuming a constant SFR within each of these bins. We then go on 
to neglect the youngest age bin, as any young stars (or hot stars, to be more precise), if present, represent a very small 
fraction of the light for the large majority of early-type galaxies \citep[90\%][]{thomas10}. 

We then compute a grid of mock galaxies over the following parameter ranges: \fehi~and \feho~vary between -0.5 and 0.2, 
\afei~and \afeo~very between -0.2 and 0.4 and the mass contribution of the intermediate stars to the total stellar mass may 
vary between 0 and 100 \%. We note that we thus do \emph{not} assume an exponentially falling star formation history at any point. 
On the other hand, for this exercise we divide the SFH of our mock galaxies in two age bins with constant star formation rate. Our grid 
of models is probably more varied than real early-type galaxies, in particular concerning the contribution of intermediate-age stars, 
and thus any conclusions we derive from here is applicable to a more restricted set of SFHs as well. All model spectra are 
convolved with a velocity dispersion of 150 km/s. We verified that our results do not depend on velocity dispersion, \emph{as 
long as this parameter is correctly determined}. One of the advantages of full spectrum fitting is that the velocity dispersion 
is determined simultaneously with the stellar population analysis, thus avoiding some of the potential pitfalls with the more 
classical Lick index method. Guided by the S/N distribution of our sample (see Figure \ref{f:ston}), we perturb the spectra 
assuming Gaussian uncertainties to the lowest S/N ratio of the sample, i.e.~40.  

\subsection{Recovery of light-weighted mean quantities}
\label{s:simul_all}

We now use {\tt paradise} to recover the following quantities: 
\begin{itemize}
\item V-band light-weighted mean parameters $\mathrm{age}$, \feh, \afe, $Z$
\item V-band light-weighted mean parameters for old component \ageo, \feho, \afeo, \Zo
\item V-band light-weighted mean parameters for intermediate component \agei, \fehi, \afei, \Zi
\end{itemize}

The mathematical equation defining these values as implemented by {\tt paradise} is 
\begin{equation}
\langle X \rangle = \frac{\sum_{i=0}^N a_i L^V_i X_i}{ \sum_{i=0}^N a_i L^V_i }, 
\label{e:lwq}
\end{equation} 
where $a_i$ are the weights for $N$ templates with properties $X_i$ as well as V-band luminosities $L^V_i$. This equation can also 
be used to define the same quantities over a restricted range of SSPs that are part of the SEH (e.g.~ those representing only old 
or only intermediate-age stars) by suitably restricting the indices over which the sum is computed. 

\begin{figure}[tbp]
\begin{center}
\resizebox{1 \hsize}{!}{\includegraphics[]{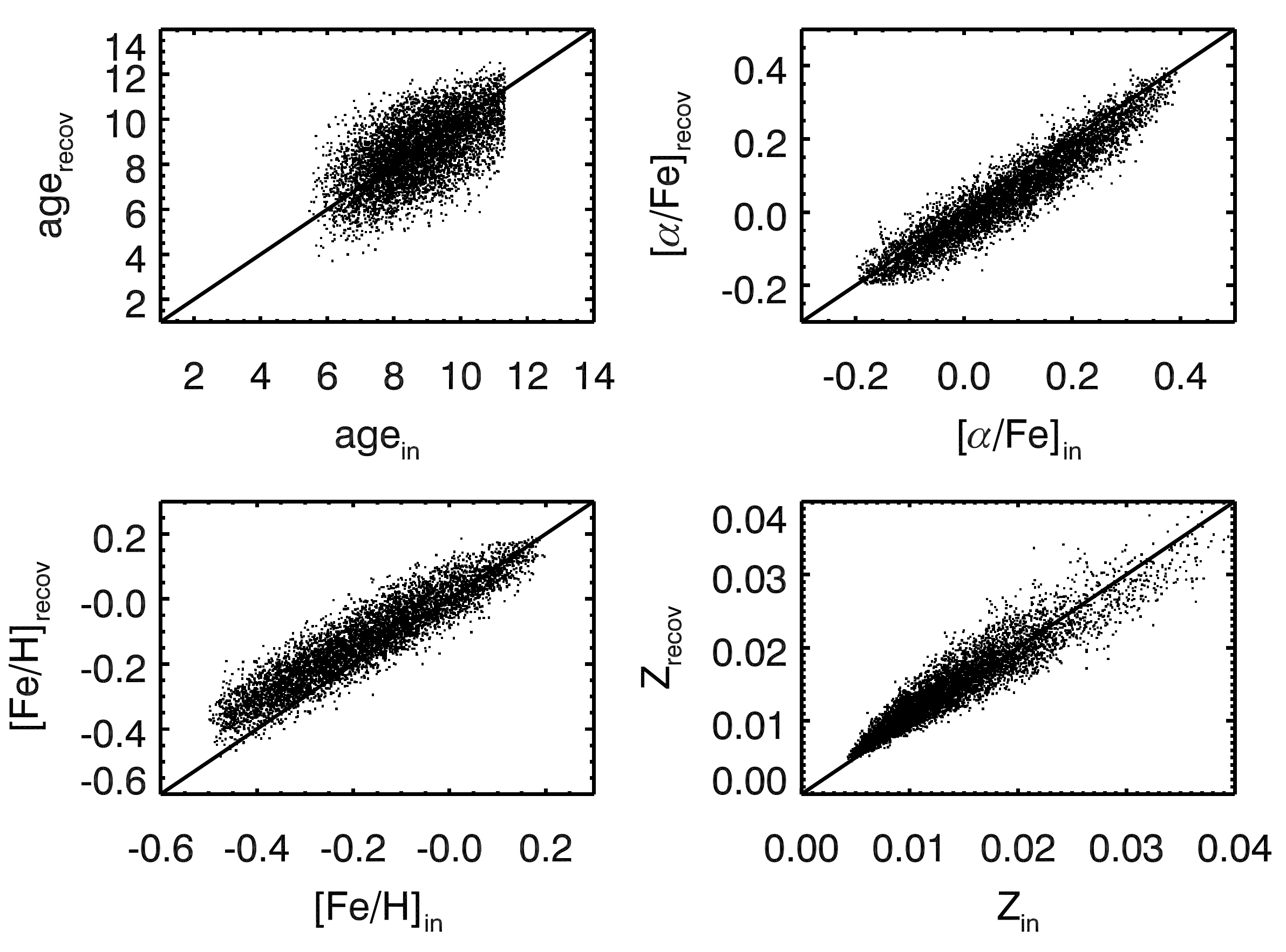}}
\end{center}
\caption[mean]{ Simulations of parameter recovery for a randomized coverage of parameter space and a total of 3000 mock galaxies.}
\label{f:mean}
\end{figure}

Our results for the overall light-weighted quantities, i.e.~when the index i in Eq.~\ref{e:lwq} runs over all templates, are summarized 
in Figure \ref{f:mean}. 

\cite{koleva08b} and \cite{sanchezblazquez11} study the influence of a metallicity -- velocity dispersion degeneracy, implying that when 
both parameters are fitted simultaneously, the velocity dispersion may be underestimated simultaneously with the metallicity or \feh. 
We find the opposite effect in our simulations, i.e.~for lower values of \feh~velocity dispersion will be over-estimated slightly (for our fiducial 
value of 150 km/s the overestimate is of order 10 km/s at \feh~= $-0.3$). We observe no 
dependence on \afe. \cite{sanchezblazquez11} emphasize that the effect is larger for young stellar populations, 
while our study here focusses on old stellar populations. Also, we are not interested in the velocity dispersions here, but in the parameters 
age, \feh~and \afe. Any influence of this metallicity -- velocity dispersion degeneracy would show up in our study as a bias in \feh~recovery. 
Indeed, this may be the reason for the slight deviation from a perfect recovery at low metallicities observed in our simulations. 

We also explicitly test the influence of either normalizing (rectifying) or not normalizing the spectra with a pseudo-continuum 
based on a running mean. We find that the fit with non-normalized continuum is formally slightly better, as expected due to the 
higher information content. We caution, however, that this 
is true for the ideal case we can test here, i.e. where the model and the "data" ideally match each other. Even a slight problem with 
flux calibration would potentially bias our results, we thus always normalize spectra from real data below. 

\begin{figure}[tbp]
\begin{center}
\resizebox{1.0\hsize}{!}{\includegraphics[]{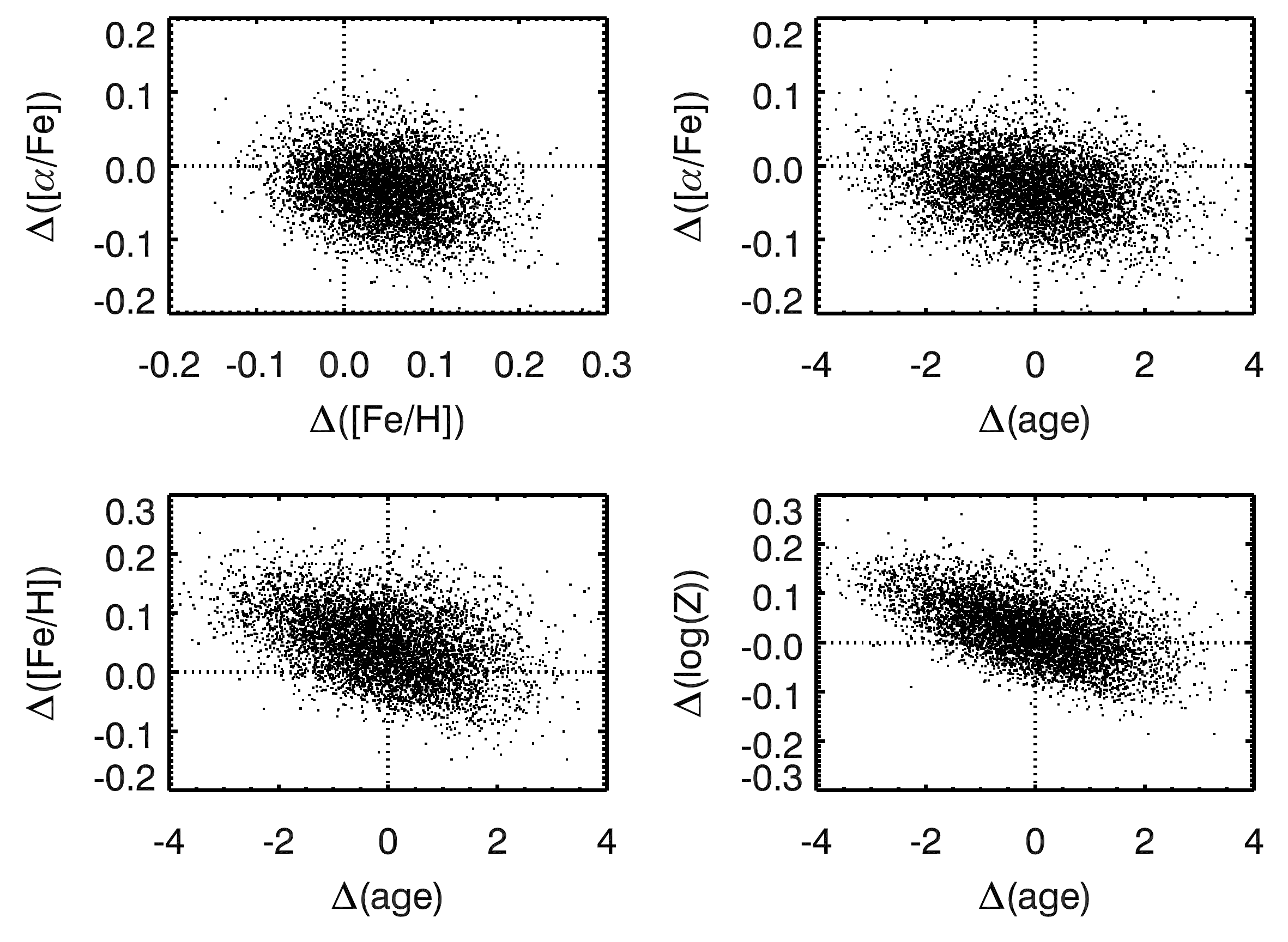}}
\end{center}
\caption[degen]{Degeneracies in the overall mean light-weighted values as difference plots. 
Each $\Delta$ is defined as the recovered minus the true value. The dominating degeneracy occurs between age and \feh, while 
age and \afe~show only weak degeneracy.}
\label{f:degen}
\end{figure}

The age-metallicity degeneracy is a long-standing problem in the study of old stellar populations. In the lower left panel in 
Figure \ref{f:degen} one can read of that the effect is of order 0.1dex in \feh~per 2 Gyr. Here we refrain from performing a 
formal fit to the simulated data. This would only pretend to provide more precise information as it is entirely unclear how the mock 
galaxy sample would relate to either the sample under study here or the real full galaxy population. 
Comparison with \citet[][their Figure 3]{thomas05}, \citet[][, their Figure 6]{cid-fernandes05}, and 
\citet[][their Figure 7]{sanchezblazquez11} shows that the trends are qualitatively the same, 
confirming once more that this effect is physical and does not depend on the fitting algorithm being applied. In particular it is 
surprisingly independent of whether the fitted stellar population is a SSP or the fitting algorithm allows for a more complex 
star formation history. To be more precise, the degeneracy is between \feh~and age. There are degeneracies between age and \afe, 
and between \feh~and \afe, but these are much less pronounced. 

A new problem in the analysis of integrated spectra from ETGs are the many lines of evidence for a variation in the initial mass 
function of galaxies \citep{van-dokkum10b,cappellari12}. 
\cite{ferre-mateu14} explore how such variations would affect the analysis of the SFHs of ETGs and 
find that complete neglect of IMF variations would potentially significantly affect the results in particular in terms of the relative 
contribution of old and intermediate-age stars. Incorporating this caveat into our analysis is at present beyond the scope of this 
contribution. 

\begin{figure}[tbp]
\begin{center}
\resizebox{1.0\hsize}{!}{\includegraphics[]{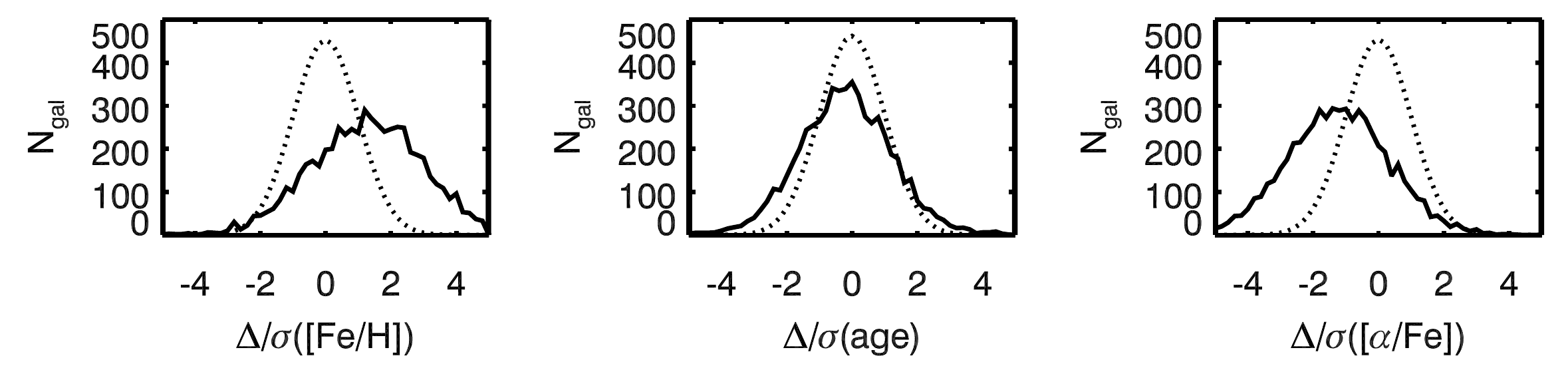}}
\end{center}
\caption[error]{Verification of the uncertainties derived by the fitting routine through bootstrap. Each $\Delta$ is defined as the 
recovered minus the true value, but this time divided by the size of the error bar (denoted $\sigma$) as output by the fitting routine. For perfectly 
Gaussian errors, the distribution should coincide with a Gaussian of width one and normalized to the same total area (dotted line).  }
\label{f:error}
\end{figure}

As described in Appendix \ref{s:paradise},  {\tt paradise} performs a bootstrap estimate of errors on the recovered stellar population 
parameters. Figure \ref{f:error} verifies that these errorbars are appropriate. Indeed, for correctly estimated Gaussian errorbars, the 
histogram of the quantity $\Delta(X)/\sigma_X$ should be a Gaussian with width 1, where 
$\Delta(X) = X_{\mathrm{measured}} - X_{\mathrm{true}}$. This is almost exactly the case for age and the widths of the distributions 
in \feh~and \afe. There are systematic offsets for \feh~and \afe. Given the average error bars from the simulations these translate to 
a systematic overestimation of \feh~by 0.07 dex and an underestimation of \afe~by 0.03 dex. Figure \ref{f:mean} shows that the largest 
offset of $\sim$0.1 dex occurs at low values of \feh. 

\subsection{Can we resolve the enrichment histories?}
\label{s:simul_split}

\begin{figure}[tbp]
\begin{center}
\resizebox{0.99\hsize}{!}{\includegraphics[]{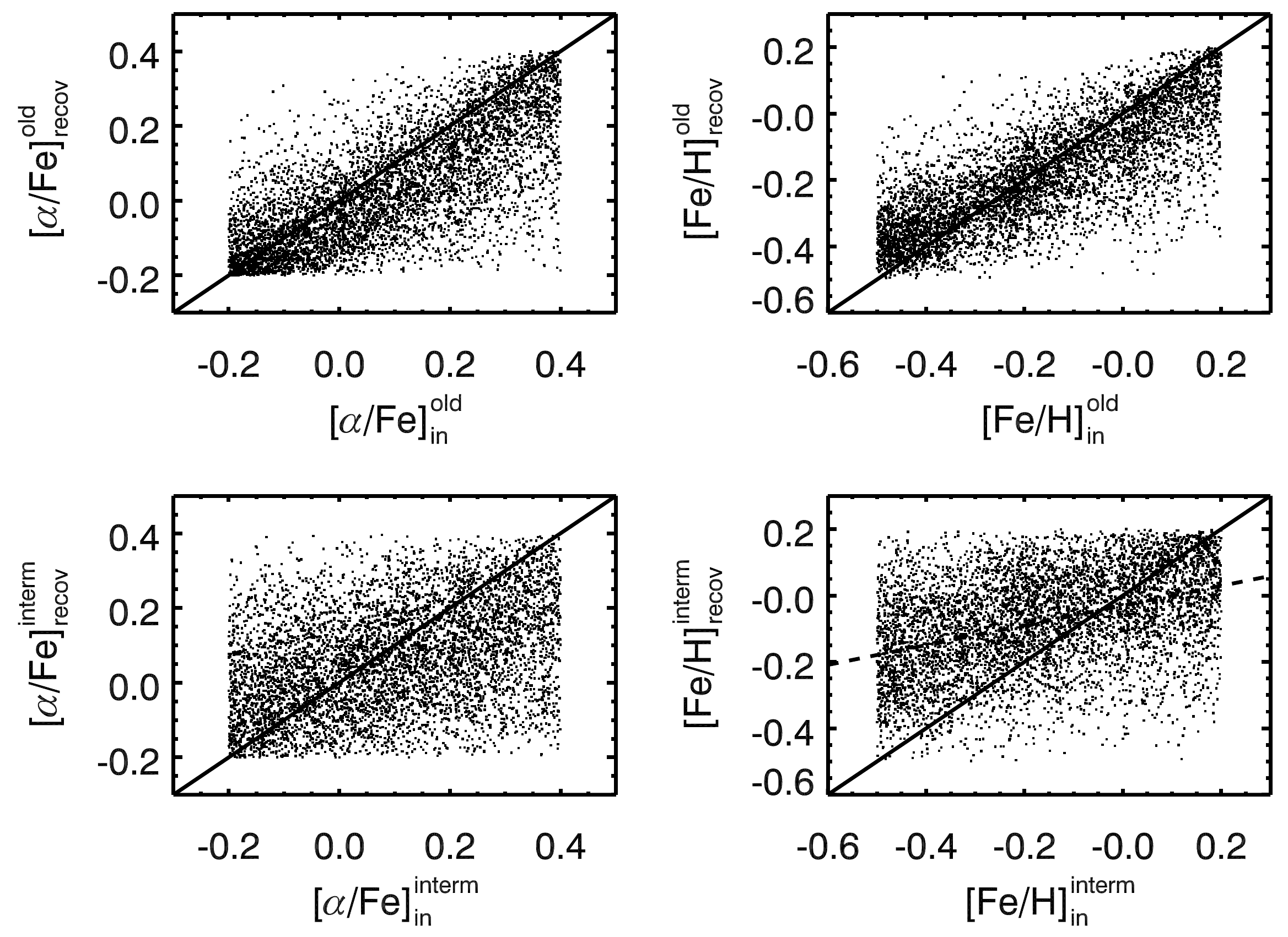}}
\end{center}
\caption[split]{Simulations of parameter recovery when splitting the star formation and enrichment history at a fiducial 
age of 9.5 Gyrs. The dashed line in the lower right panel is a formal fit to the points in the data with a slope of $\sim0.3$.} 
\label{f:split}
\end{figure}

The term SEH implies the wish to represent the relation between age and abundance for each galaxy. To our knowledge this 
has not been done in the literature. While \cite{de-la-rosa11} have applied the spectral fitting technique to the SFH, they 
have not actually resolved the SEH, preferring to quote a single value for the enhancement in $\alpha$-elements. We thus here 
investigate in detail, whether our 
method is able to produce this kind of information reliably. We thus repeat the same plots of input vs.~recovered quantities as in 
Section \ref{s:simul_all} for the properties \feho, \afeo, \fehi, \afei. The result is shown in Figure \ref{f:split}. To economize some space, 
we do not plot ages and metallicities for the old and intermediate distinction. The reason is that the ages are fixed already by the 
way we separate the star formation history into bins and its recovered distribution therefore carries no real meaning (although it 
can vary slightly within the width of that bin). Metallicity on the other hand is only a trivial combination of \feh~and \afe. 

Figure \ref{f:split} looks encouraging concerning the old stars in that clear relations scattering around the one-to-one line exist, 
with an RMS width of about 0.1 dex. While it may not seem intuitive at first sight that the properties of the older stellar population, 
which contributes on average less to the total light than the younger or intermediate-age population, would be well-determined 
from a fit to the integrated spectrum, this is actually in line with the findings of \cite{serra07}. These authors found that the 
SSP-equivalent chemical composition depends mainly on the chemical composition of the older of two constituting stellar 
populations. It is also in line with theoretical expectations that the effects of \afe~(in particular Mg) are stronger for cooler stars 
\citep[\feh~][]{cassisi04, coelho07, sansom13}.

\begin{figure}[tbp]
\begin{center}
\resizebox{0.99\hsize}{!}{\includegraphics[]{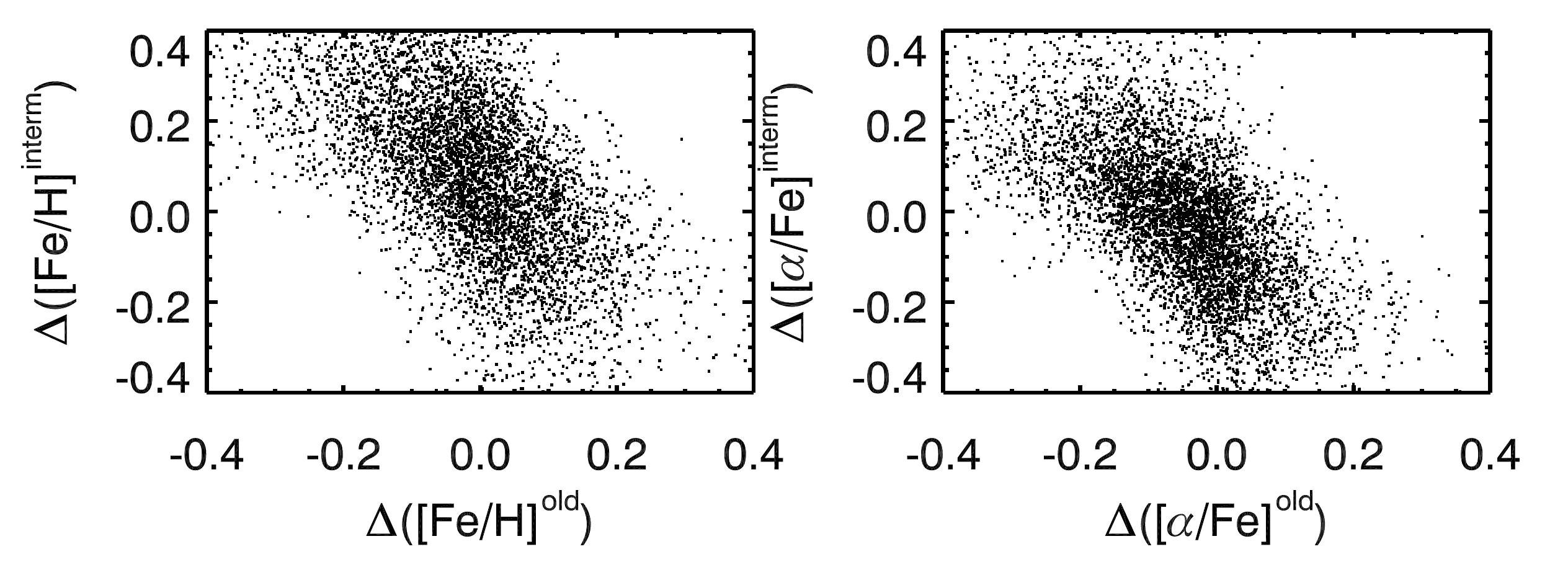}}
\end{center}
\caption[splityo]{The dominating degeneracies in the mean light-weighted values when separating SEHs into intermediate and 
old stars are those between the same properties of the two stellar generations. Each $\Delta$ is defined as the recovered minus 
the true value. } 
\label{f:degenyo}
\end{figure}

This would in turn predict that the properties of the younger population are less well traced. Indeed, 
in the case of \fehi~and \afei~the slope of the relation is clearly lower than one. A formal 
fit (the dashed line) yields slopes of $\sim 0.3$. To investigate the origin of this slope we have also investigated degeneracies 
between different quantities after this split into old and intermediate-age population. The same plots as in Figure \ref{f:degen} for 
the separation of the SEH into old and intermediate populations shows a much larger scatter, but no noticeable increase in the 
degeneracies, thus these plots are not shown here in the interest of space. However, and as shown in Figure \ref{f:degenyo}, 
the dominating degeneracies are those between \fehi~ and \feho~ and between \afei~ and \afeo.

At this point we remind the reader of the classification into SLF and ACC galaxies introduced in Section \ref{s:intro}. Indeed, at specific 
points in this paper we separate galaxies in two types: 1) Those where \feho~$<$ \fehi. For these the \feh~evolution with time is 
consistent with self-enrichment by SNe Ia and these are therefore termed SLF-ETGs (represented by yellow dots in Figs. 17 to 
22 below). 2) Those galaxies where \feho~$>$ \fehi.  Those are not consistent with self-enrichment and we infer that some 
accretion of material must have taken place -- whether these were mergers, gas accretion or anything else we remain agnostic 
about \citep[such as winds redistributing elements within a galaxy, see][]{pipino06}. Nevertheless, for conciseness we 
term those merger-accretion or ACC-ETGs (pink color in same figures). A more in-depth discussion of the physical usefulness 
of this distinction is given in Section \ref{s:mergers}. 

We now explore how robust such 
a classification would be to degeneracies. Our input mock galaxy catalogue contains 50\% SLF and 50\% ACC galaxies. After fitting, 
81\% of the SLF galaxies would be correctly identified but only 33\% of the ACC galaxies. Furthermore, after fitting we would identify 74\% 
of the sample as SLF galaxies and 26\% as ACC galaxies. Finally, the number that is probably of most interest is the percentage of 
galaxies in each half of the parameters space that actually is what it pretends to be: The percentage of galaxies that would be classified 
as SLF galaxies after fitting and that really are SLF galaxies is 55\%, while the percentage of galaxies that would be classified as ACC 
after fitting and that are really ACC galaxies is 65\%. While these numbers are sobering, one should remember that they apply to a 
sample of mock galaxies that completely fills the parameter space available to our models and at a specified S/N of 40. 
This does \emph{not} represent the real 
distribution of galaxies in this parameter space as we shall see below. In particular we shall explore in an observationally motivated 
way how our conclusions depend on the exact location of the separation \feho~$<$ \fehi~+ $\Delta$, i.e. how this classification 
behaves within a range of $-0.4 < \Delta < 0.4$. 

\begin{figure*}[tbp]
\begin{center}
\resizebox{0.49\hsize}{!}{\includegraphics[]{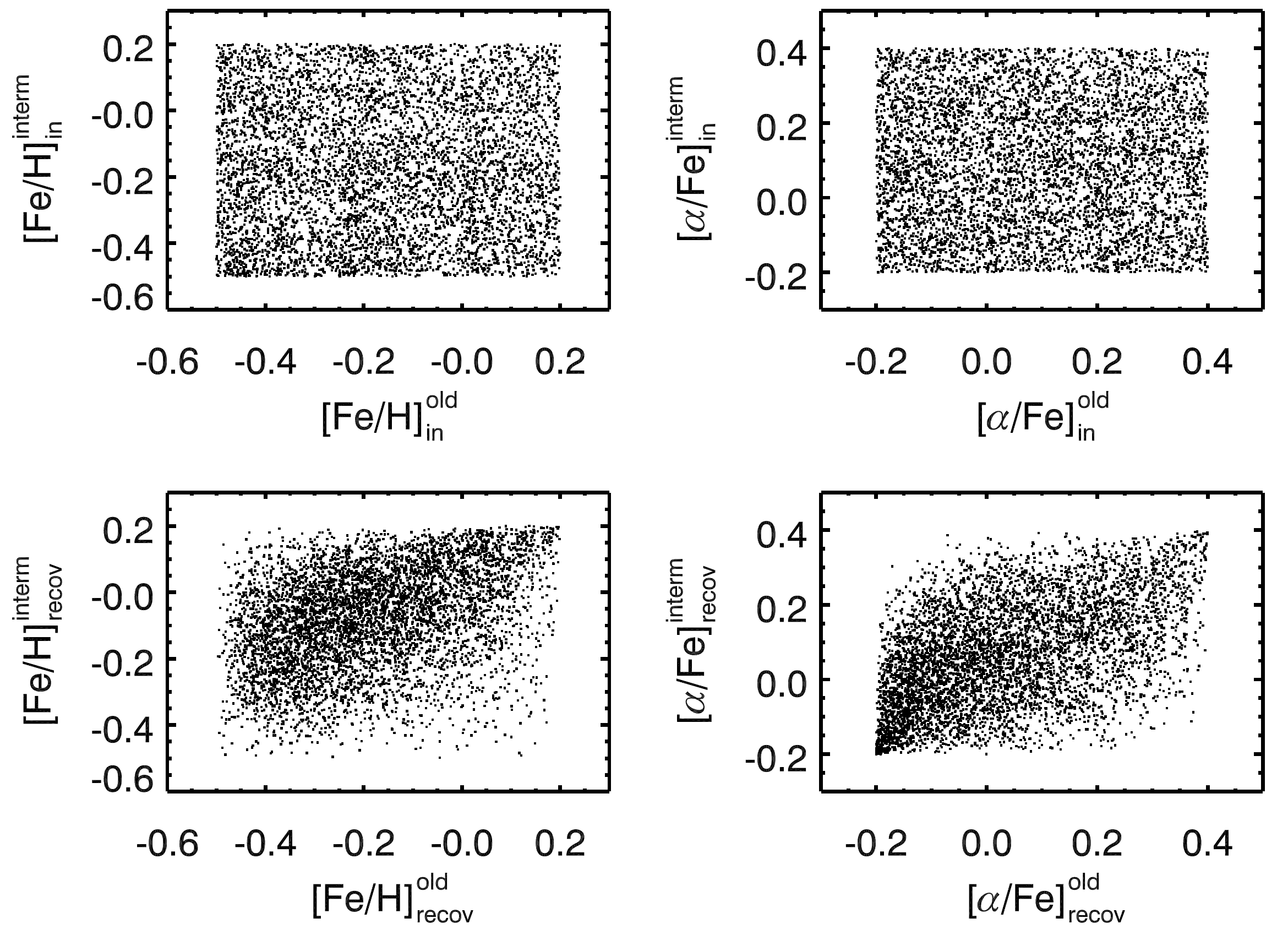}}
\resizebox{0.49\hsize}{!}{\includegraphics[]{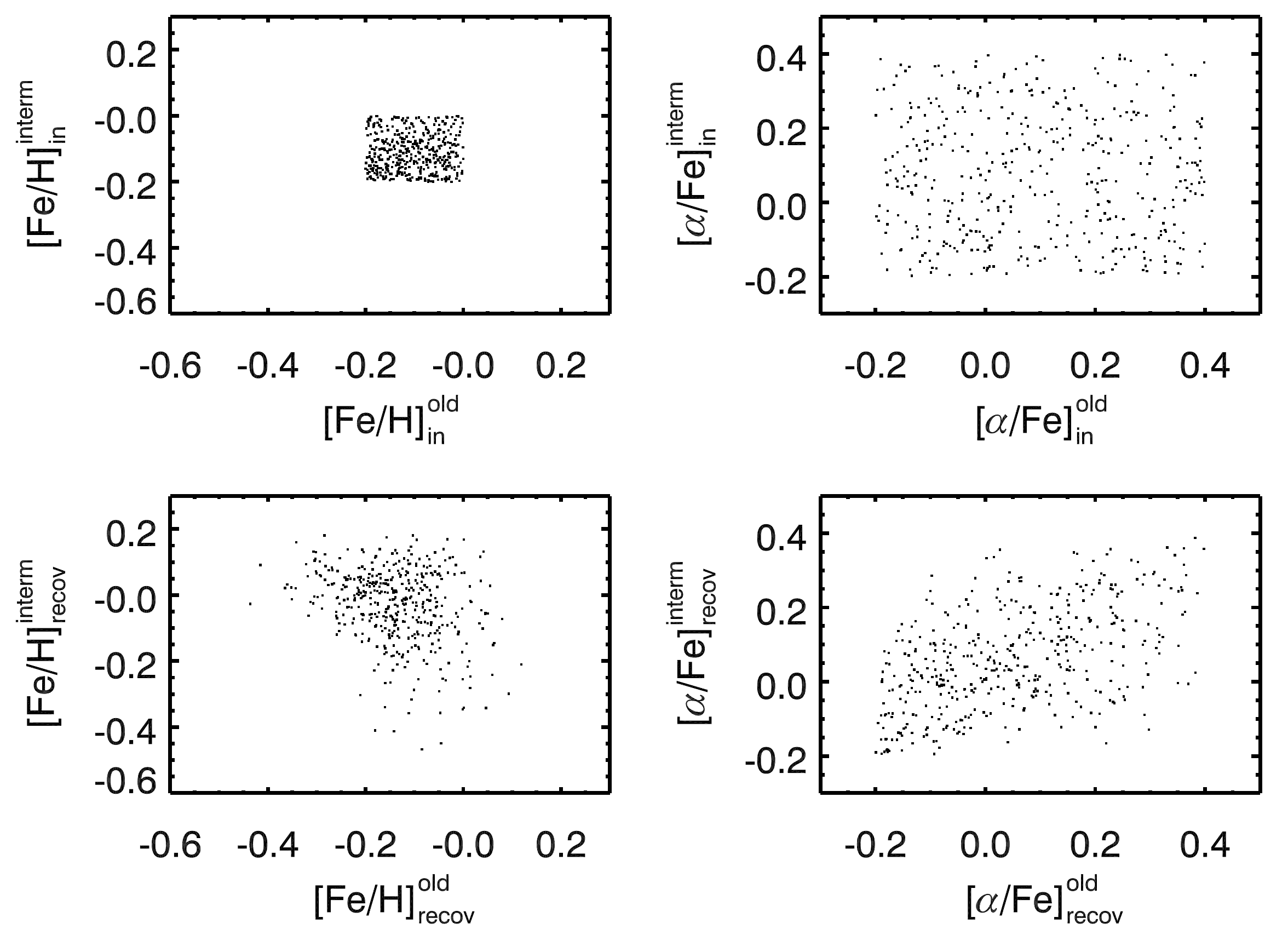}}
\end{center}
\caption[distrib]{Comparison of the original distribution of parameters (upper row) vs.~the one recovered after the fitting process 
(lower row). We show the two parameter spaces that are of most interest in the remainder of the paper, namely \fehi~vs.~\feho~ 
and \afei~vs.~\afeo. \emph{Left panel:} The case in which the total library of mock galaxies is analyzed.  \emph{Right panel:} The 
case in which only mock galaxies with $-0.2 \langle$ \fehi, \feho~$\langle 0.0$ are analyzed. } 
\label{f:distrib}
\end{figure*}

Besides direct classification we use another, more statistically minded method to gauge the distribution of observed galaxies 
in the planes of \fehi~vs.~\feho~and \afei~vs.~\afeo, namely the posterior distribution of fitted galaxies. We show the general case 
in Figure \ref{f:distrib}: one starting with a uniform distribution of mock galaxies in parameter space. For reasons to be seen later, 
another interesting case is the one where all galaxies start out in a narrow range of \fehi~and \feho~abundances. We have chosen 
$-0.2 <$ \fehi, \feho~$< 0.0$. Clearly both cases produce significantly different posterior distributions in parameter 
space that can be used to constrain the real distribution of parameters to some extent -- or at least can be used to exclude some 
real parameter distributions. 

In this section we have assessed the uncertainties intrinsic to our method \emph{under the assumption that the model is a 
perfect representation of the data.} We have found that these uncertainties exist, but that we can 
understand, describe and quantify them. 

\section{Fitting the observed spectra}
\label{s:fit1}

We now proceed to fit the sample described in Section \ref{s:sample} with the method described in Section \ref{s:simul}. 
We did not change the setup of the fitting in any way. This means that we fit the spectra with the same models, the same 
continuum rectification, and all other parameters being equal. There is only a slight difference in the spectra that may affect 
the fitting procedure to some extent.

\begin{figure}[tbp]
\begin{center}
\resizebox{1.0\hsize}{!}{\includegraphics[]{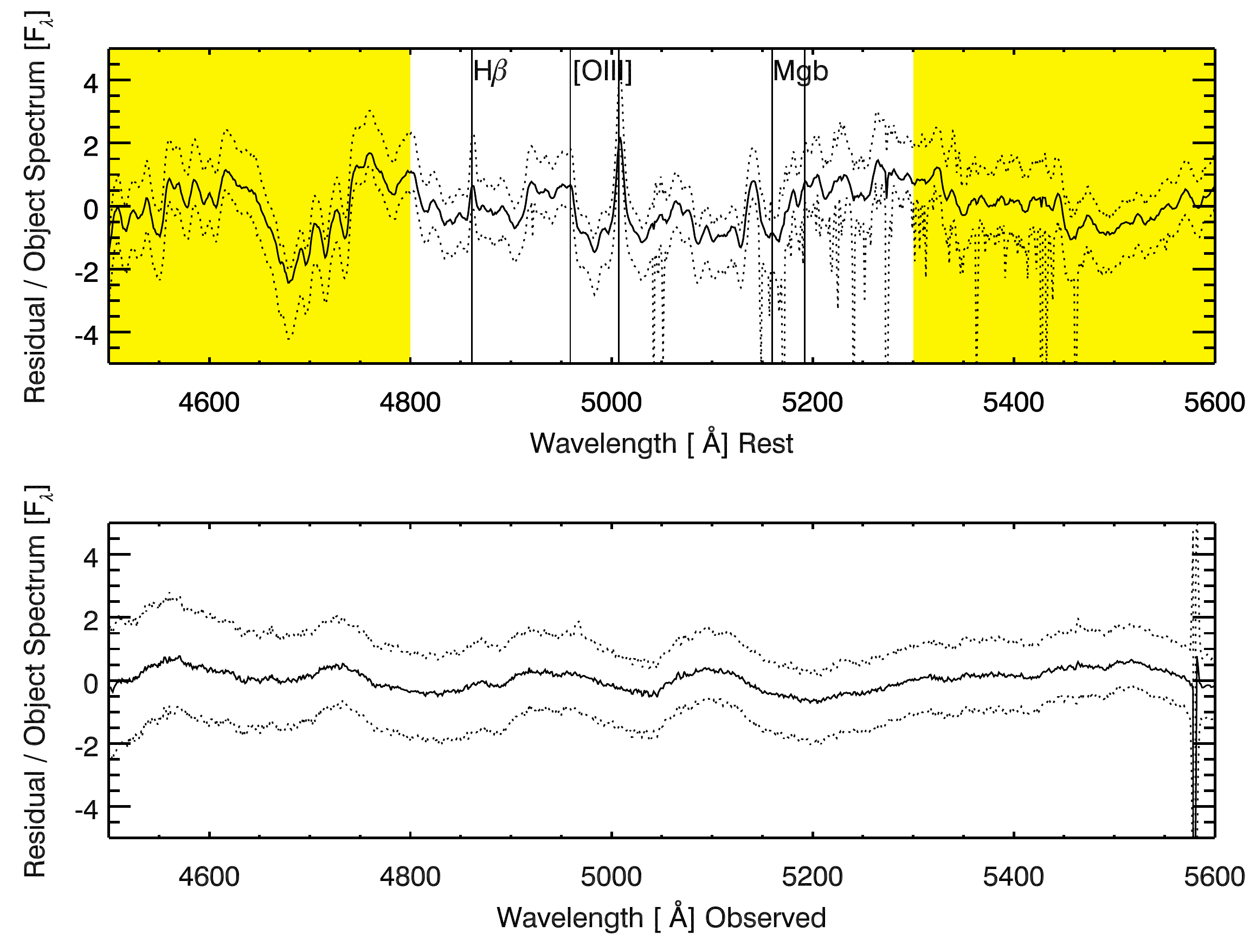}}
\end{center}
\caption[residuals]{Mean residual vector and its standard deviation from all galaxies in the sample. The upper plot indicates 
the residuals in the rest frame of the galaxies, the yellow region has actually not been used to determine the best fit. [OIII] and 
H$\beta$ emission lines are labelled and clearly visible. The Mgb feature is also indicated for reference, but does not show 
evidence of specific mismatch. 
The lower plot shows the observed frame residuals indicating no major concern with the data. }
\label{f:residual}
\end{figure}

This is seen in the stacked fitting residual vectors of all galaxies as shown in Figure \ref{f:residual}. Our sample has been 
selected to contain emission-line free galaxies as judged by \cite{brinchmann04}. However, we find that the residuals do contain 
emission lines of H$\beta$ and [OIII]. We attribute this finding to our excellent stellar population modeling and fitting procedure 
as well as to conservative estimates of emission line absence. In other words, while to the trained eye emission line residuals in 
stacked spectra may seem obvious, for an automatic detection procedure it is difficult to attribute a statistical significance 
to these features \emph{per galaxy}, given similar-sized residuals in nearby regions of the spectrum. Emission lines affect only 
small regions of the spectrum and thus should not affect our overall results. In many cases they are masked away by our automatic 
outlier rejection in {\tt paradise} (see Appendix \ref{s:paradise}). We have verified that all our simulation results are 
independent of the potential missing information in the emission line region by applying an appropriate mask in wavelength space 
to the mock galaxies and rerunning all simulations. We found no difference in the results. 

As a side-note we mention the residual absorption feature around 4700 {\AA}. Its origin is unclear, but it does not affect our measurements 
as it is outside our fitted wavelength range. 

We also investigated the presence of any problems with data quality. 
Given the precision we are claiming to reach here (not much more than one percent in 
residual) we might be prone to data quality problems at a level not often seen otherwise. Problems might be present in the SDSS 
data, but also in the stellar templates. We note, however, that Figure \ref{f:residual} does not indicate such problems. Indeed, the lower panel 
shows residuals in the observed frame where data problems such as sky emission line residuals should appear. The only obvious 
sky residual is at the very red end of the wavelength range shown, but way outside our fitted wavelength range. A paper 
dealing in much more detail with template mismatch vs. data quality is in preparation. 

\begin{figure}[tbp]
\begin{center}
\resizebox{1.0\hsize}{!}{\includegraphics[]{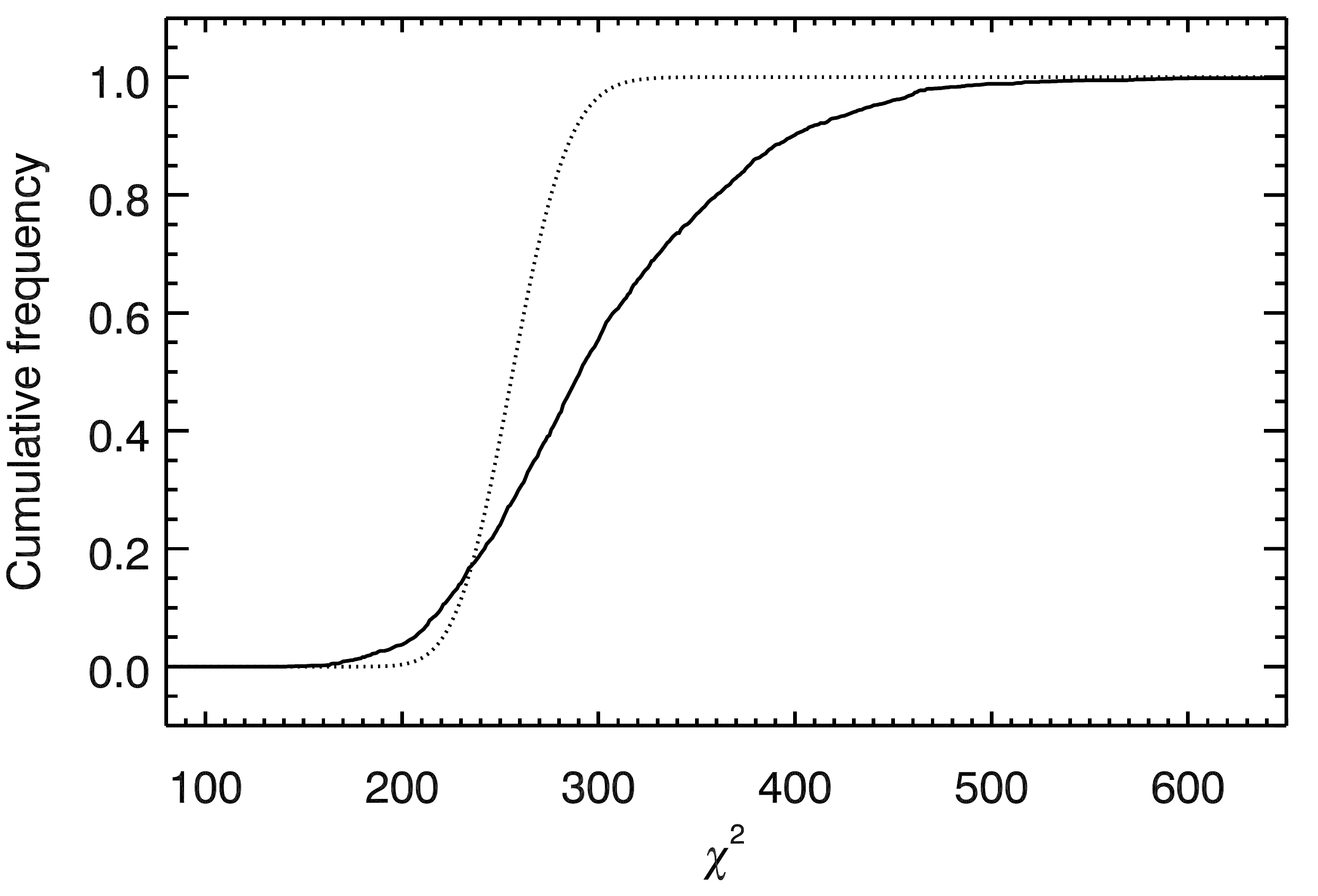}}
\end{center}
\caption[chisq_distrib]{Cumulative histogram of $\chi^2$ values obtained from the data (solid line) with the $\chi^2$ distribution 
expected from statistics (dotted line) for 257 degrees of freedom. }
\label{f:chi2distrib}
\end{figure}

Another way to assess our fit quality is to compare the cumulative histogram of $\chi^2$ obtained from the spectra in our data 
with the one expected from statistics, i.e. the $\chi^2$-distribution. This is done in Figure \ref{f:chi2distrib}, assuming 257 
degrees of freedom for our fitting 
procedure, as appropriate after taking into account the number of fitted pixels and the number of free parameters. The distributions do 
not match perfectly, but they are clearly very similar, both in shape and location. We believe this indicates that the fitting of SDSS spectra 
is indeed in the realm governed by $\chi^2$ statistics, once the models are sophisticated enough and when restricting to a well-known 
wavelength range. 

\begin{figure}[tbp]
\begin{center}
\resizebox{1.0\hsize}{!}{\includegraphics[]{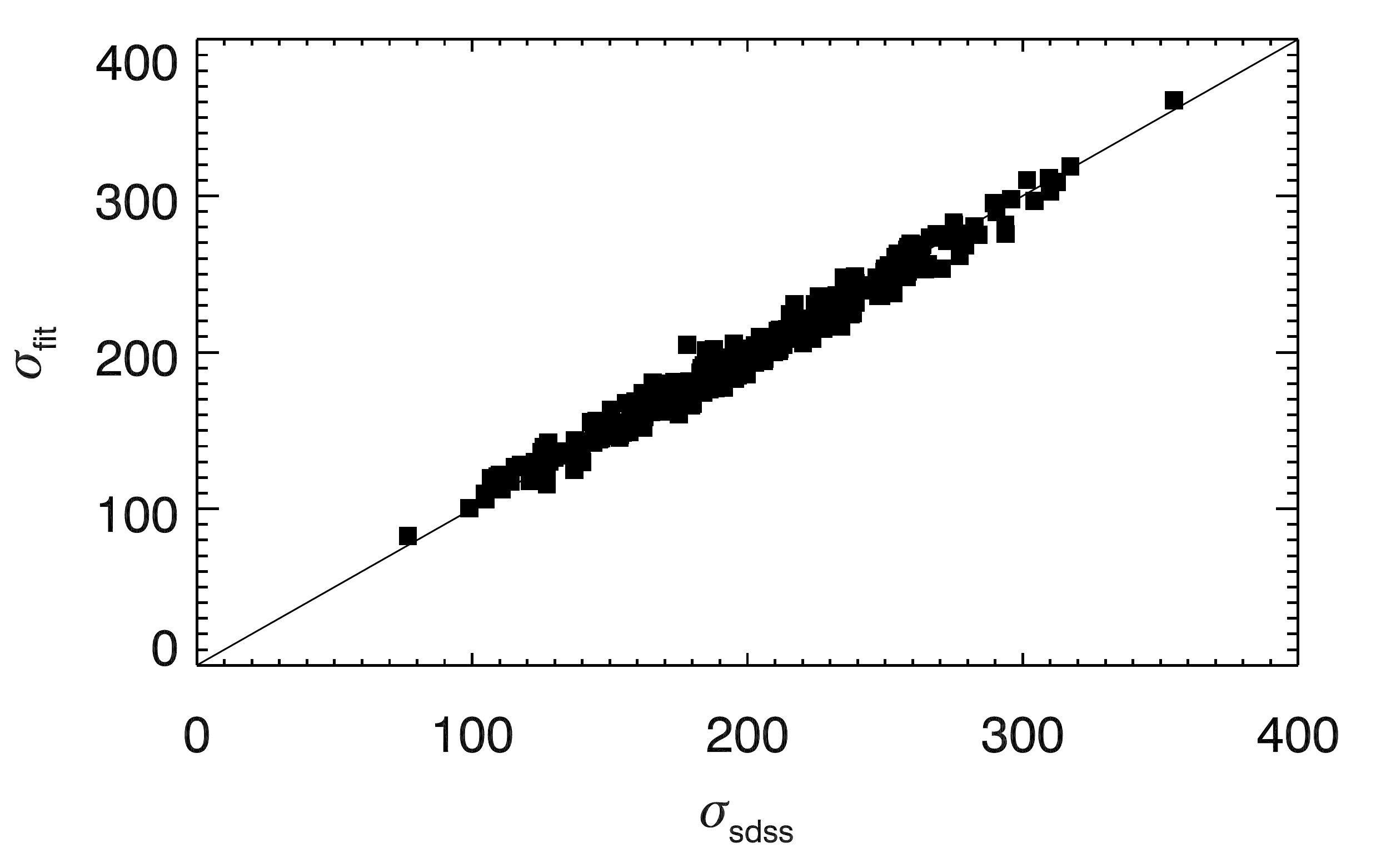}}
\end{center}
\caption[residuals]{Comparison between the velocity dispersion from SDSS DR7 and those obtained in this work.}
\label{f:veldisp}
\end{figure}

In Section \ref{s:simul} we used a velocity dispersion of 150 km/s for all tests, although the velocity dispersion was left a free parameter 
in the fitting. Velocity dispersion is clearly different for every observed galaxy. We can, however, feed the SDSS pipeline velocity dispersion 
as a starting value to {\tt paradise}, thereby essentially mimicking the fact that we know the velocity dispersion for the simulated galaxies. 
Indeed, we find that we do recover the velocity dispersion from the SDSS very well, see Figure \ref{f:veldisp}.\footnote{There is one object 
out of 2286 where the SDSS pipeline velocity dispersion is 224.8 km/s, whereas {\tt paradise} determines 118.9 km/s. Visually this object 
clearly has been correctly fitted by {\tt paradise}. We only have this one example, which however tentatively indicates that velocity 
dispersion mismatch would not have been a problem for our analysis. }

The output of the fitting are tables with all parameters, such as mean light-weighted age (\feh, \afe) of the entire spectrum, 
the same parameters only for the "old" stars (i.e. older than 9.5 Gyrs) and for the "intermediate-age" stars (i.e. younger than 
9.5 Gyrs). Typical errorbars on our parameters are 0.2 Gyr on age and 0.01 dex on \afe~and \feh. These errorbars reflect random 
errors and degeneracies, but not model-related errors and systematic problems with the recovery. This can be compared 
with the typical errors quoted in one of the classical studies of the field. For definiteness we choose \cite{thomas05}, without 
prejudice to later or earlier results. They quote typical errors of 1.48 Gyr in age, 0.04 dex in total metallicity, and 0.02 dex in \feh~ratio, 
although it needs to be said for clarity that these earlier errorbars only take into account the formal uncertainties on the data. Our errorbars 
also include degeneracies in the stellar population mix. In the more recent paper by \cite{johansson12} the uncertainty in [O/Fe] at S/N $\sim$ 
40 in the $r$-band is given as approximately 0.05 dex, but it needs to be kept in mind that this is the uncertainty on a single element, 
while we group all $\alpha$-elements together. 

Another important point to mention is that none of the derived galaxy ages are older than 13 Gyr, as we show in Section \ref{s:results}. 
While this is predetermined from the range of model spectra in use, there is also no hint of an edge effect in the 
sense the galaxies that would be older than 13 Gyr cannot be fitted and therefore for those galaxies a higher metallicity would 
be fitted to compensate. We explicitly tested that this edge effect is present if we restrict our template library to SSPs with ages 
younger than 13Gyr. 

\section{Results}
\label{s:results}

We now discuss our findings in terms of correlations between the parameters we determined ourselves, i.e.~ages, \feh~and \afe~and 
those available for our galaxies i.e.~stellar mass M* and velocity dispersion $\sigma$. 

\subsection{The $\sigma$-$\alpha$, M*-$\alpha$ and age-$\alpha$ relations}

\begin{figure*}[tbp]
\begin{center}
\resizebox{1.0\hsize}{!}{\includegraphics[]{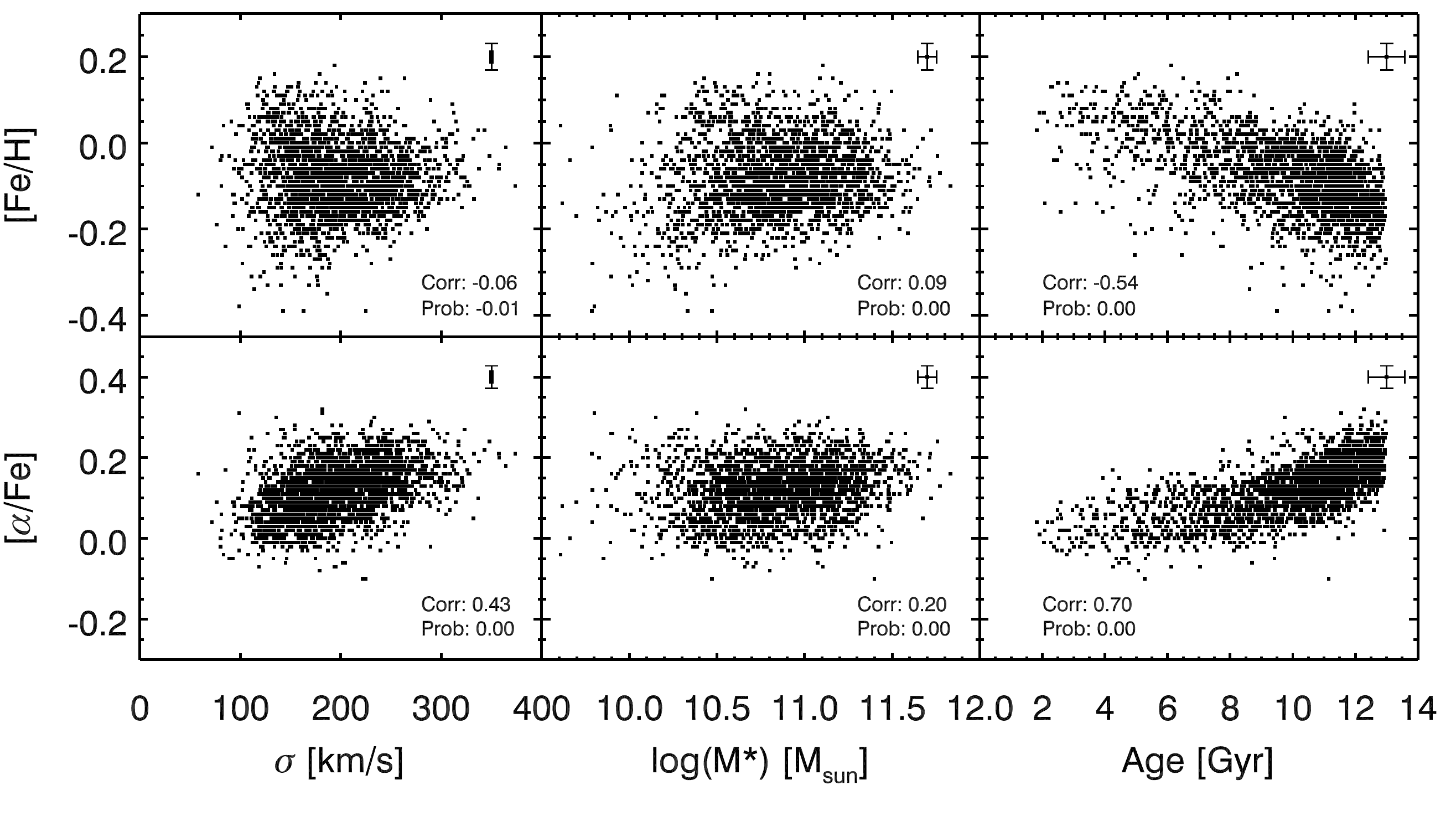}}
\end{center}
\caption[mean]{Correlations of \feh~and \afe~ with velocity dispersion, stellar mass and mean light-weighted age for our sample 
galaxies. It is clear that the tightest relations are those with age. Note also the change of slope in the relation between age and 
\afe$\sim$9Gyr. Average error bars are given in the upper right corner of each panel and spearman rank correlation coefficients 
are also noted. }
\label{f:mean_quant}
\end{figure*}

In Figure \ref{f:mean_quant} we plot \feh~and \afe~versus velocity dispersion $\sigma$, stellar mass M* and mean light-weighted age. 
We quantify the presence or absence of correlations by means of Spearman rank coefficient tests\footnote{The Spearman coefficients 
measure the strength of a potential correlation, where 0 is no correlation, 1 is a strong positive correlation and -1 is a strong anti-correlation. 
The probabilities on the other hand measure the significance of the correlation, i.e. the probability that the real correlation is zero. Therefore 
a value of 1 for this probability indicates that we know nothing about the real correlation, while a value of below 0.05 would imply that a 
correlation indeed exists at the the 95\% confidence level.} and by means of formal fits to each set of two parameters combinations 
using the {\tt LINFIT} module in IDL. The results of the fits are given in Table \ref{t:overall}. 
We confirm correlations between log($\sigma$) as well as log(M*/\msun) and \afe~as found in previous work (see Section \ref{s:earlier}). 
In Figure \ref{f:mean_quant} we also see a strong correlation 
between age and \afe~as well as \feh~in the sense that \afe~decreases for younger light weighted mean age of the galaxy, while \feh~
increases. 

There seem to be only weak correlations between \feh~and log($\sigma$). However, \cite{johansson12} show that when the sample 
is decomposed into age bins, each age bin by itself shows a strong correlation with log($\sigma$). In the interest of space we refrain 
from repeating the exercise. 

\begin{table}
\centering
\caption{Correlation strengths for light-weighted mean parameters }
\begin{tabular}{c|cc}
Parameters & Intercept  & Slope \\
\hline
log($\sigma$) vs \feh&  0.02$\pm$0.01 & -0.05$\pm$0.005 \\
log($\sigma$) vs \afe& -0.46$\pm$0.01 &  0.26$\pm$0.005 \\
log(M*/\msun) vs \feh& -0.44$\pm$0.02 &  0.03$\pm$0.002 \\
log(M*/\msun) vs \afe& -0.20$\pm$0.02 &  0.03$\pm$0.002 \\
age vs \feh&  0.15$\pm$0.003 & -0.02$\pm$0.001 \\
age vs \afe& -0.09$\pm$0.002 &  0.02$\pm$0.001 \\
\end{tabular}
\label{t:overall}
\end{table}

Additionally the relation between age and \afe~shows a clear change of slope at age $\sim$9 Gyr. We have fitted the two regimes 
separately and find that \afe~= -0.01$\pm$0.004 + 0.01$\pm$0.001$\cdot$age for galaxies younger than 9 Gyr, while 
\afe~= -0.20$\pm$0.007 + 0.03$\pm$0.001$\cdot$age for galaxies older than 9 Gyr. For both regimes the probability of the absence of any correlation is zero.  

\begin{figure}[tbp]
\begin{center} 
\resizebox{1.0\hsize}{!}{\includegraphics[]{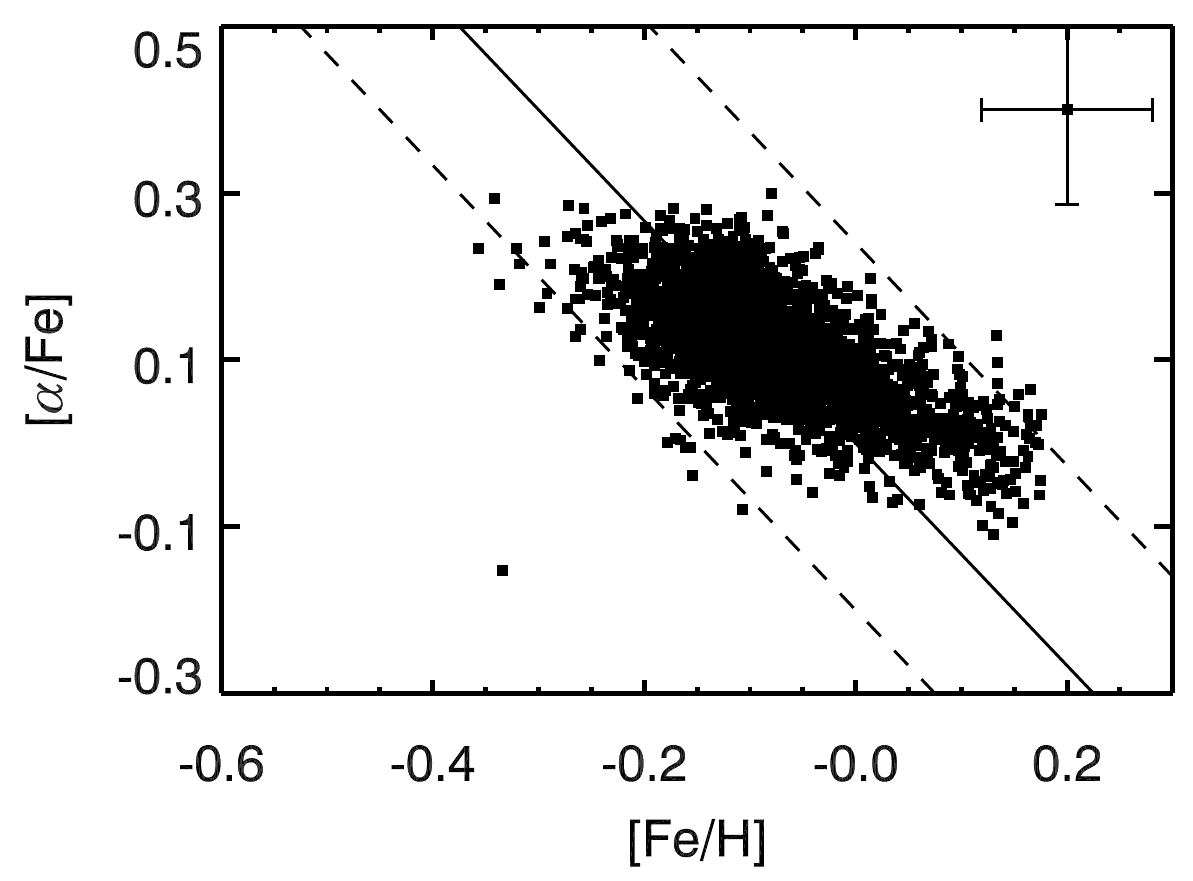}}
\end{center}
\caption[mean]{Correlation between \feh~vs. \afe~for the luminosity weighted average properties of the ETGs in our sample. 
The thick solid line is a line of constant, solar 
metallicity (Z=0.017). The two dashed line illustrate the extremes of the ETG locus, i.e. for [Z]=-0.15 and [Z]=0.18. }
\label{f:metallicity}
\end{figure}

The two rightmost panels of Figure \ref{f:mean_quant} already indicate a correlation between \afe~and \feh, which we explicitly plot in Figure 
\ref{f:metallicity}. We find that early-type galaxies all have on average a total metallicity that is roughly solar. It is the ratio between the abundances of Fe and the $\alpha$-elements that changes between galaxies. 

\subsection{Separating the old and intermediate components}
\label{s:separate}

We now show results for resolved SEHs, i.e. looking at the abundance ratios of old and intermediate stellar populations separately. 
Before we do so, as a cautionary remark, it is worth emphasizing that the uncertainty bars include the uncertainties introduced by the degeneracies characterizing galaxy evolution. Thus a direct visual comparison of the size of the error bar and the scatter in the figures is 
misleading. The size of the error bar indeed denotes the range of possible values for each single galaxy. However, on average, 
galaxies will tend to be assigned the mean value of that possible range. Scatter in figures is mainly 
driven by real observational uncertainties and does not represent the degeneracy uncertainties as assessed by {\tt paradise}. 
That said, mean uncertainties on our parameters are 0.17 dex for \fehi, 0.14 dex for \feho, 0.12 for \afei, and 0.13 for \afeo. 

\begin{figure}[h]
\begin{center}
\resizebox{1.0\hsize}{!}{\includegraphics[]{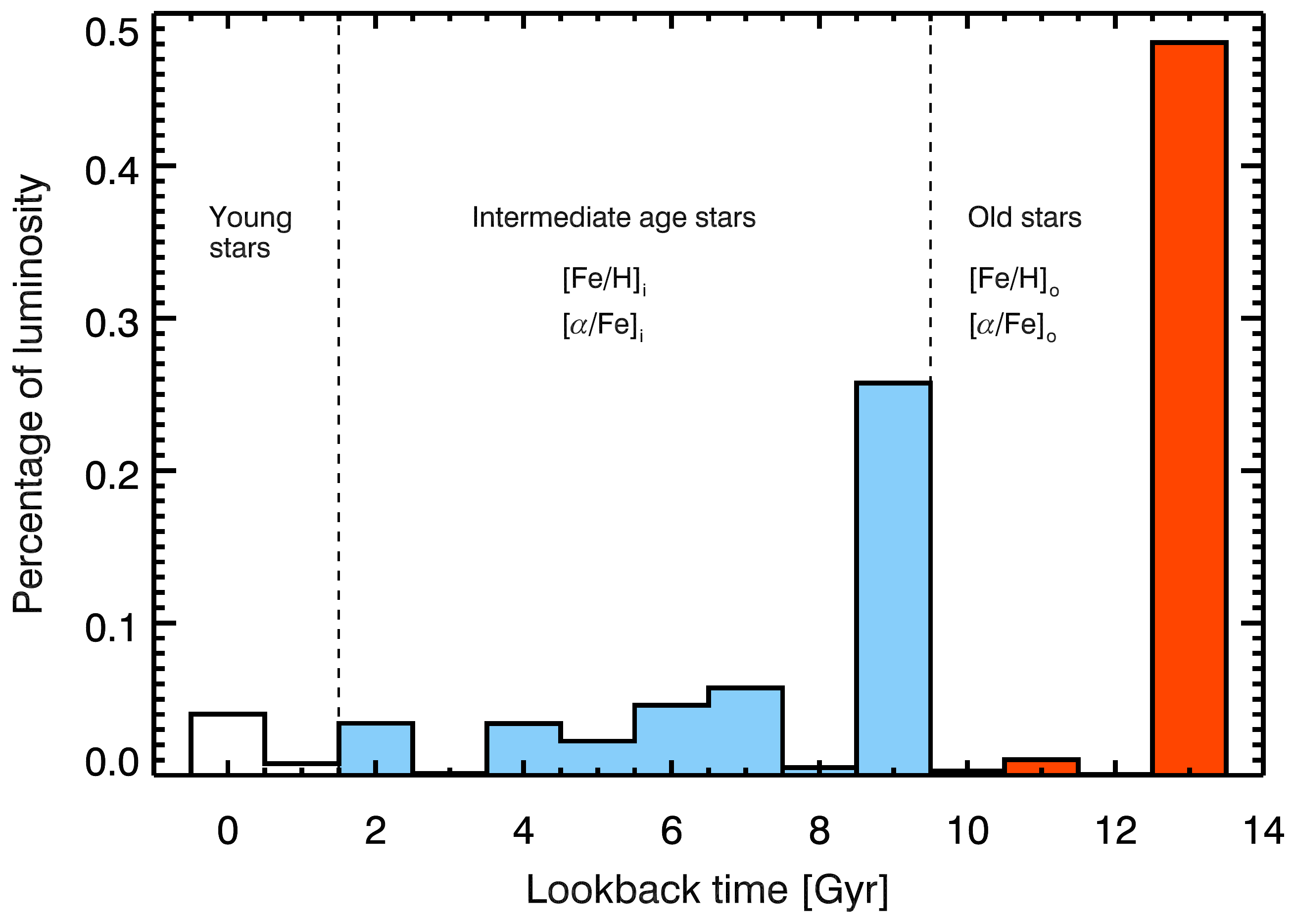}}
\end{center}
\caption[sfh]{Average star formation history of all galaxies in the observed sample in terms of the present day contribution of 
each stellar population to the total luminosity of the galaxy. All galaxies contribute equally to this average, 
i.e. the total luminosity for each galaxy has been normalized to one before averaging. }
\label{f:sfh}
\end{figure}

First we repeat in Figure \ref{f:sfh} the upper panel of Figure \ref{f:didactic}, i.e. we show the non-binned average star formation 
history of all galaxies in our sample. As this is the observable, we show the star formation history in terms of the luminosity contributions 
of stars of each specified age to the total luminosity of each galaxy. It is noteworthy that several peaks in the star formation history 
appear, namely at 13 Gyr, 9 Gyr and a somewhat broader one around 6 Gyr. Our SSPs are sampled in steps of 1Gyr, therefore intermediate 
ages could in principle be assigned significant weights by {\tt paradise}. On one hand it is entirely possible that this is an artifact 
of our fitting procedure, given that such features do not appear in the overall star formation history of the universe (Lilly-Madau plot). 
On the other hand, the two older episodes of star formation seem entirely compatible with the classical two-infall scenario of chemical 
evolution \citep{chiappini97} that was originally proposed for the Milky Way. Note that these two oldest peaks also contribute to 
justifying our separation into old and intermediate-age stars at a lookback time of 9.5 Gyr. We come back to this point below 
when discussing SLF vs. ACC-ETG percentages.

Finally, there is a small peak at less than 1 Gyr which is outside the coverage of our \afe-enhanced SSP templates. As described 
in Section \ref{s:models} we include five templates with younger ages, but with solar element ratios only. The total contribution to 
the light of the population younger than 2 Gyr is however less than 5\%. Due to its lower mass to light ratio than the older stellar 
populations this translates to an even smaller contribution to the total mass of the galaxy. We thus confirm that ETGs harbor 
very little young stars -- maybe even none, if currently unmodeled old and hot stellar populations indeed do play a role 
\citep[see e.g.][]{ocvirk10}. 

\begin{figure*}[tb]
\begin{center}
\resizebox{1\hsize}{!}{\includegraphics[]{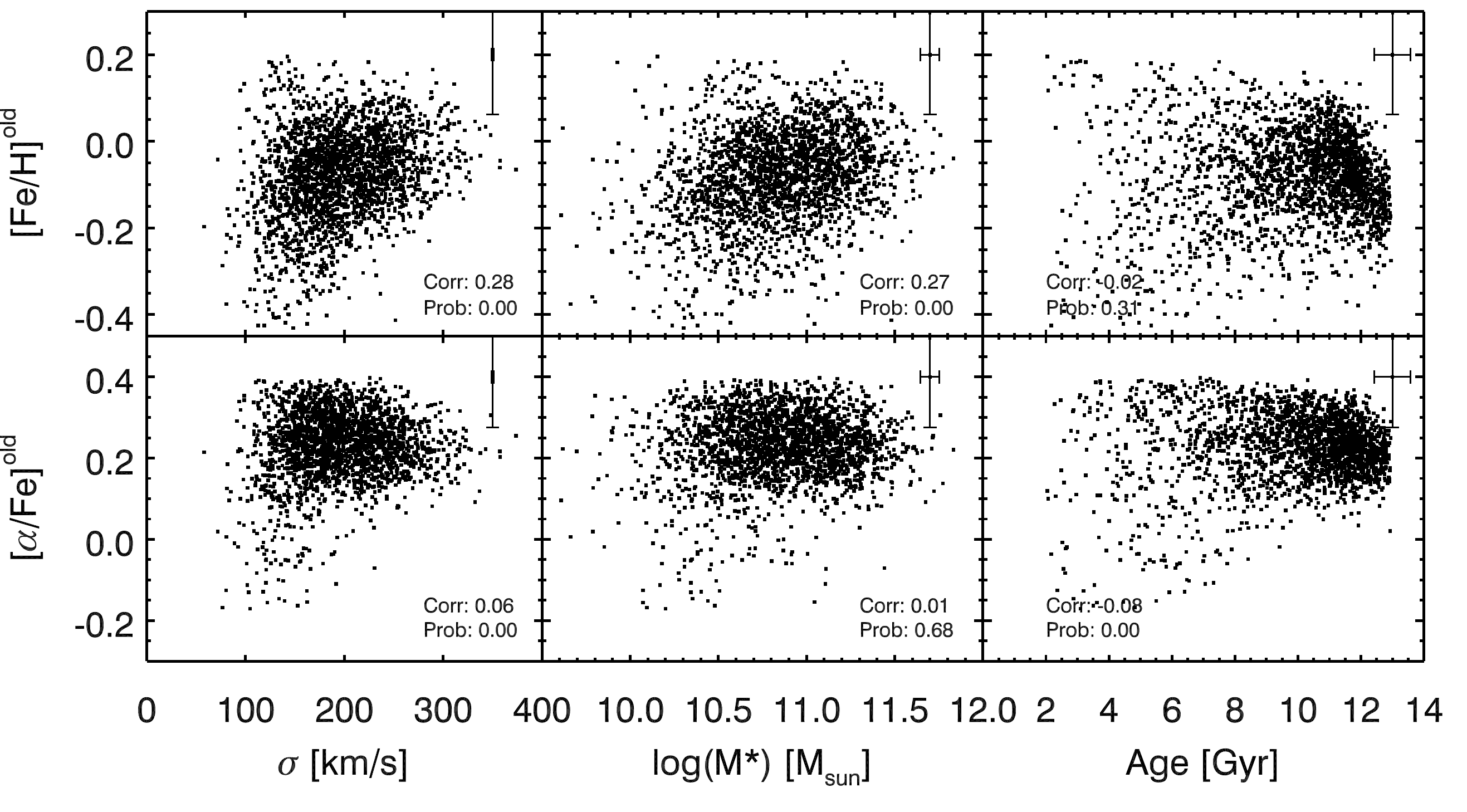}}
\end{center}
\caption[oy_vs_all]{Correlations of the \feho~and \afeo~with the global properties of the sample galaxies $\sigma$, M*, and age. 
Average error bars are given in the upper right corner of each panel and spearman rank correlation coefficients are also noted. }
\label{f:agemassig_old}
\end{figure*}

\begin{table}
\caption{Correlation strengths old stars only }
\begin{tabular}{c|cc}
Parameters & Intercept  & Slope \\ 
\hline
log($\sigma$) vs \feho&  -0.73$\pm$0.046 & 0.29$\pm$0.020 \\
log($\sigma$) vs \afeo&  0.37$\pm$0.046 & -0.05$\pm$0.020 \\
log(M*/\msun) vs \feho& -0.83$\pm$0.066 &  0.07$\pm$0.006 \\
log(M*/\msun) vs \afeo&  0.46$\pm$0.063 & -0.02$\pm$0.006 \\
age vs \feho&  0.13$\pm$0.013 & -0.02$\pm$0.001 \\
age vs \afeo&  0.39$\pm$0.012 & -0.01$\pm$0.001 \\
\end{tabular}
\label{t:old}
\end{table}

It is interesting to start the analysis with plotting the properties of the old stellar populations vs. the global parameters of their host 
galaxies. We refrain from doing so for the intermediate-age populations due to the strong degeneracies we identified in 
Section \ref{s:simul_split} and the lower recovery success for the intermediate component with respect to the old one we identified there. 
Also, our sample is dominated by galaxies in which the old stellar population contributes more than half 
of the total light, contributing to possible uncertainties in the properties of the intermediate-age stellar populations. 
Figure \ref{f:agemassig_old} shows that the correlations seen for the global stellar populations do not 
hold up for the older stars in the galaxies (compare also Table \ref{t:old}). Indeed, while the Spearman rank test indicates a 
very low probability for the 
dependence of \afeo~and \feho~on age to be spurious, inspection of the plot tells another story. There is a cloud of points 
at older ages, while the galaxies with younger ages show much larger scatter. This can be understood as being 
caused by difficulties in accurately estimating \afeo~and \feho~for galaxies with significant intermediate-age populations. 
We therefore attach no importance to these relations. For $\sigma$ and M* vs. \afeo,  
the Spearman test indicates weak correlation and indicates a high probability (0.68) of no correlation for M* vs. \afeo. We are 
thus left with the indication that \feho~depends positively on mass as parameterized by either $\sigma$ or M*, but more strongly 
on the former. This correlation is driven mainly by the presence of objects with low \feho~at low $\sigma$ values. Overall these 
trends are opposite to the average properties, where \afe~depends clearly on mass and \feh~does not. 

\begin{figure}[h]
\begin{center}
\resizebox{1.0\hsize}{!}{\includegraphics[]{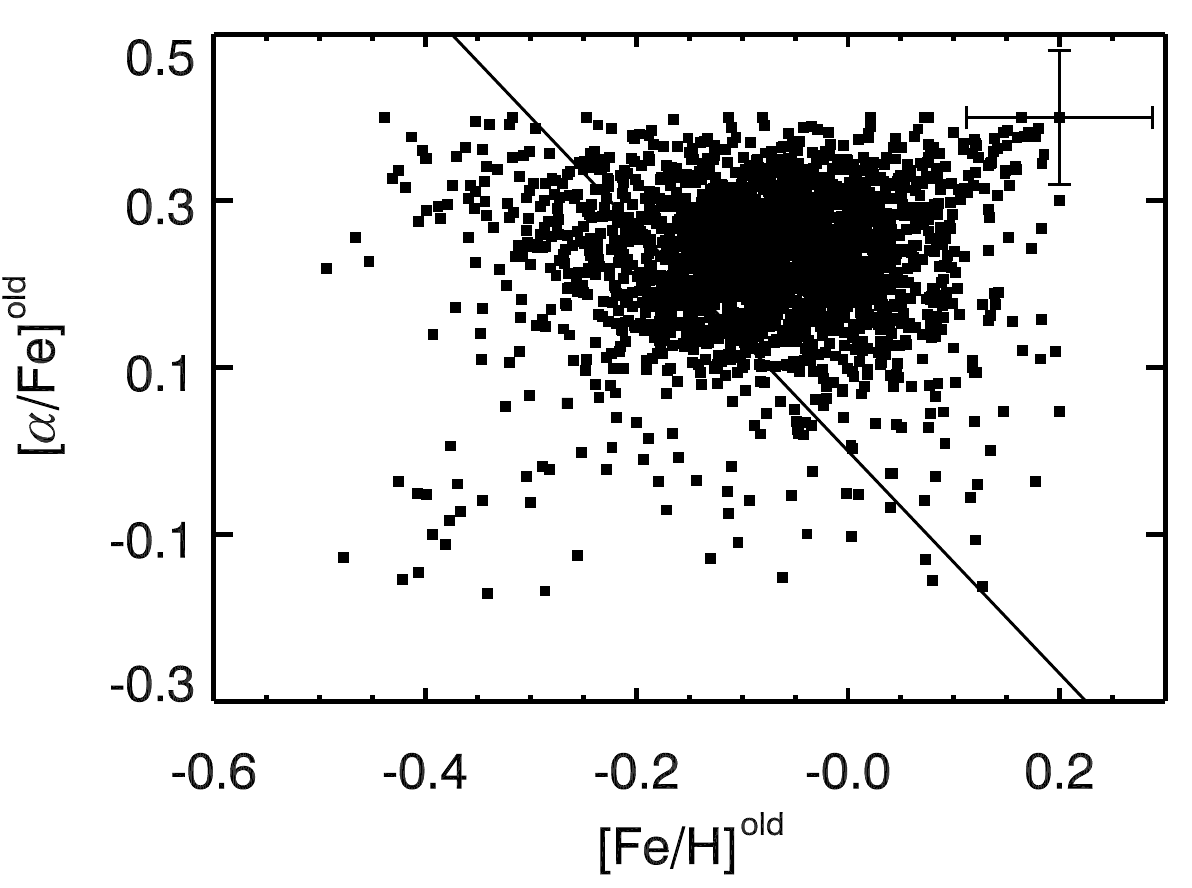}}
\end{center}
\caption[mean]{Relation between \feho~vs. \afeo, the thick line is a line of constant, solar metallicity (Z=0.017). This is the 
same as Figure \ref{f:metallicity}, but only for the stars older than 9.5 Gyr in each galaxy. The old stars in early-type 
galaxies, have on average high total metallicity, significantly super-solar. Average error bars are shown in the upper right 
corner.  }
\label{f:metallicity_old}
\end{figure}

The two rightmost panels of Figure \ref{f:agemassig_old} already indicate that the correlation between \afe~and \feh~that is valid 
on average for early-type galaxies does not hold up when considering the old stars only. Indeed, in Figure \ref{f:metallicity_old} 
we find that old stars in early-type 
galaxies have on average super-solar metallicity, where both \feho~and \afeo~are super-solar. That these properties seem to be 
independent of mass (see Figure \ref{f:agemassig_old}) might be one of the most puzzling findings of this paper. 

\subsection{Exploring the possible dichotomy SLF vs. ACC} 
\label{s:dichotomy}

\begin{figure*}[tbp]
\begin{center}
\resizebox{1\hsize}{!}{\includegraphics[]{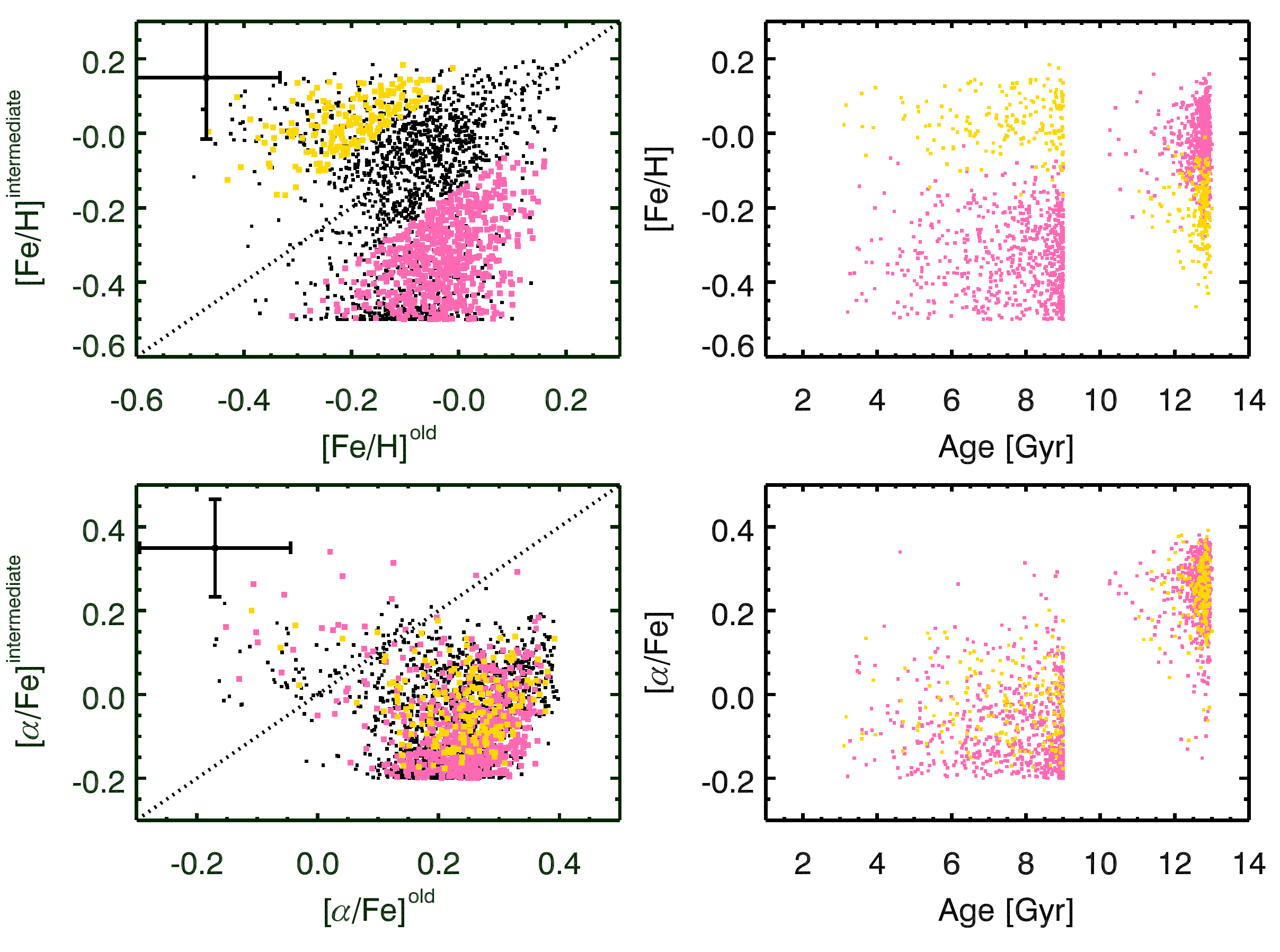}}
\end{center}
\caption[distinction]{\emph{Left panels:} The distribution of mean light-weighted \feh~and \afe~for the old vs. intermediate 
sub-populations. Galaxies in yellow are those where \feho~$<$ \fehi-0.15~(SLF-ETGs), pink where \feho~$>$ 
\fehi+0.15~(ACC-ETGs). Galaxies with uncertain classification are shown in black. Average uncertainties including all 
degeneracies are shown in the upper left corners. \emph{Right panels:} The age dependence of \feh~and \afe~within 
each galaxy, i.e.~their enrichment history. Galaxies with uncertain classification have been omitted for clarity. } 
\label{f:distinction}
\end{figure*}

In Figure \ref{f:distinction} we show the distribution of \feh~and \afe~of the intermediate vs. the old component (left panels). 
In \feh~galaxies can lie both above the identity line and below it. In \afe, almost all galaxies lie below the identity line, with a few 
slight outliers that can be thought of as being due to uncertainties. When looking back at Figure \ref{f:distrib} and the discussion in 
Section \ref{s:simul_split}, it is obvious that any statement on the distribution in these panels must take into account the uncertainties 
due to the degeneracies in properties between old and intermediate-age stars. In turns out it is easier to discuss the results in terms 
of what the data are \emph{incompatible with} and this is what we do. 

The distribution of points in the upper left panel on Figure \ref{f:distinction} is incompatible with a flat distribution in that plane. I.e. 
early-type galaxies do not have a random distribution of \fehi~and \feho. Rather their distribution is more similar to the right panels 
in Figure \ref{f:distrib}, where the underlying sample had a narrow distribution of \fehi~and \feho~centered around -0.1 for both. 

The distribution of points in the lower left panel of Figure \ref{f:distinction} is equally incompatible with a flat distribution in that plane. I.e. 
early-type galaxies do not show a random distribution of \afei~and \afeo. It seems rather secure that the oldest generation of stars has 
a high \afe~ratio, higher than 0.1, whereas the intermediate-age stars have a lower ratio of \afe, certainly below 0.1 and going 
as low as our model permits, i.e.~-0.2. Note that \afe~values below zero for late stellar generations are a generic prediction of 
chemical evolution models in massive galaxies \citep[\feh~][their Fig.~5]{minchev13}. On the other hand low-mass galaxies 
do not necessarily show low \afe. \cite{recchi01} for example show that stellar winds tend to carry elements produced 
by SNeIa away, thus leaving behind gas with higher \afe~than expected from the extent of their SFH. 

Coming back to the distinction discussed in Section \ref{s:simul_split} between ETGs that are compatible with self-enrichment 
(SLF-ETGs) and mergers/accretion (ACC-ETGs), we wish to identify galaxies that we can reasonable safely assign to any of these 
categories. SLF-ETGs are those where  \feho~$<$ \fehi~- $\Delta$, and ACC-ETGs are those where \feho~$>$ \fehi~+ $\Delta$. 
Motivated by the sizes of the error bars quoted in the beginning of Section \ref{s:separate} and after some testing in the simulations, 
for definiteness we choose $\Delta = 0.15$. Our results qualitatively do not depend on the exact size of $\Delta$. \footnote{For this 
$\Delta$ and \emph{for the simulations}, the percentage of galaxies identified as ACC-ETGs after the fitting process, which are truly 
ACC-ETGs is 77\%. The percentage of galaxies identified as SLF-ETGs after the fitting process, which are truly SLF-ETGs is 63\%.} 
We emphasize that this choice of $\Delta$ also mitigates concerns over the subtle drop in metallicity at late times shown in 
the paper by \citet{vazdekis96} and discussed in the introduction. 
We call the remaining galaxies that are neither one nor the other the 'gray-ETGs'. Note that for the observational results we include 
into the class of 'gray' galaxies those galaxies that have less than 15\% of light in either the old or intermediate stellar populations. 
This is to account for possible uncertainties in stellar population modeling and in the fitting algorithm. 

For the sample of 2286 galaxies analyzed here we find that 29\% belong to the ACC-ETG class. 63\% are unclassifiable, but 
most of those would be called ACC-ETGs as well if not for the requirement to have 15\% of light at least in the intermediate-age 
population. Only 8\% of the galaxies are bona-fide SLF-ETGs, i.e. consistent with a self-enriching star formation history. 

We caution that effects of degeneracies between the properties of the intermediate and old stellar populations, sample selection 
effects, the uncertainty on the exact boundary to be applied between old and intermediate stellar populations, specific effects 
of chemical evolution such as the drop in metallicity for young stars in massive ETGs, all contribute to make these numbers 
indicative only. To indicate the possible size of changes due to at least one of these effects we have recomputed the percentages of 
SLF and ACC-ETGs for a choice of 8.5 Gyr as the boundary between old and intermediate-age ETGs. We find that this inverts their 
relative importance, i.e.~30\% of the sample would then be SLF ETGs, while only 3\% bona-fide ACC-ETGs remain. This high sensitivity 
of the results to the adopted age boundary is due to a very low \feh~abundance attributed by the fitting code to the spike in the SFH 
at 9 Gyr. This significantly lowers the average \feho~abundance, thus increasing the number of SLF-ETGs. 

Given the magnitude of this effect, we have to carefully state our arguments for choosing 9.5 Gyr lookback time as our boundary. 
Indeed, we have argued in the introduction on physical grounds that 9.5 Gyr is a more physical boundary. On technical grounds, we 
refer to Figure \ref{f:sfh} which shows that the 9 Gyr age bin in the finely sampled SEH has a close significant neighboring age bin 
at younger age (7 Gyr), while the next significant old age bin (at 13 Gyr) is more distant in look-back time. Generally speaking, it 
is clear that there are significant degeneracies between the properties of adjacent bins in the reconstructed SEH. We have shown 
the extent of these degeneracies in Figure \ref{f:degenyo} for the coarsest possible choice of SEH, namely an SEH with two age bins. 
Degeneracies only increase for an SEH with a finer age grid. Degeneracies between the 7 and 9 Gyr age bins shown in 
Figure \ref{f:sfh} are therefore much more likely than between the 
9 and 13 Gyr age bins. This is the technical reason why we believe the results based on a 9.5 Gyr boundary to be more reliable 
and significant than those we would have obtained based on any other boundary. 

\begin{figure*}[tbp]
\begin{center}
\resizebox{1.0\hsize}{!}{\includegraphics[]{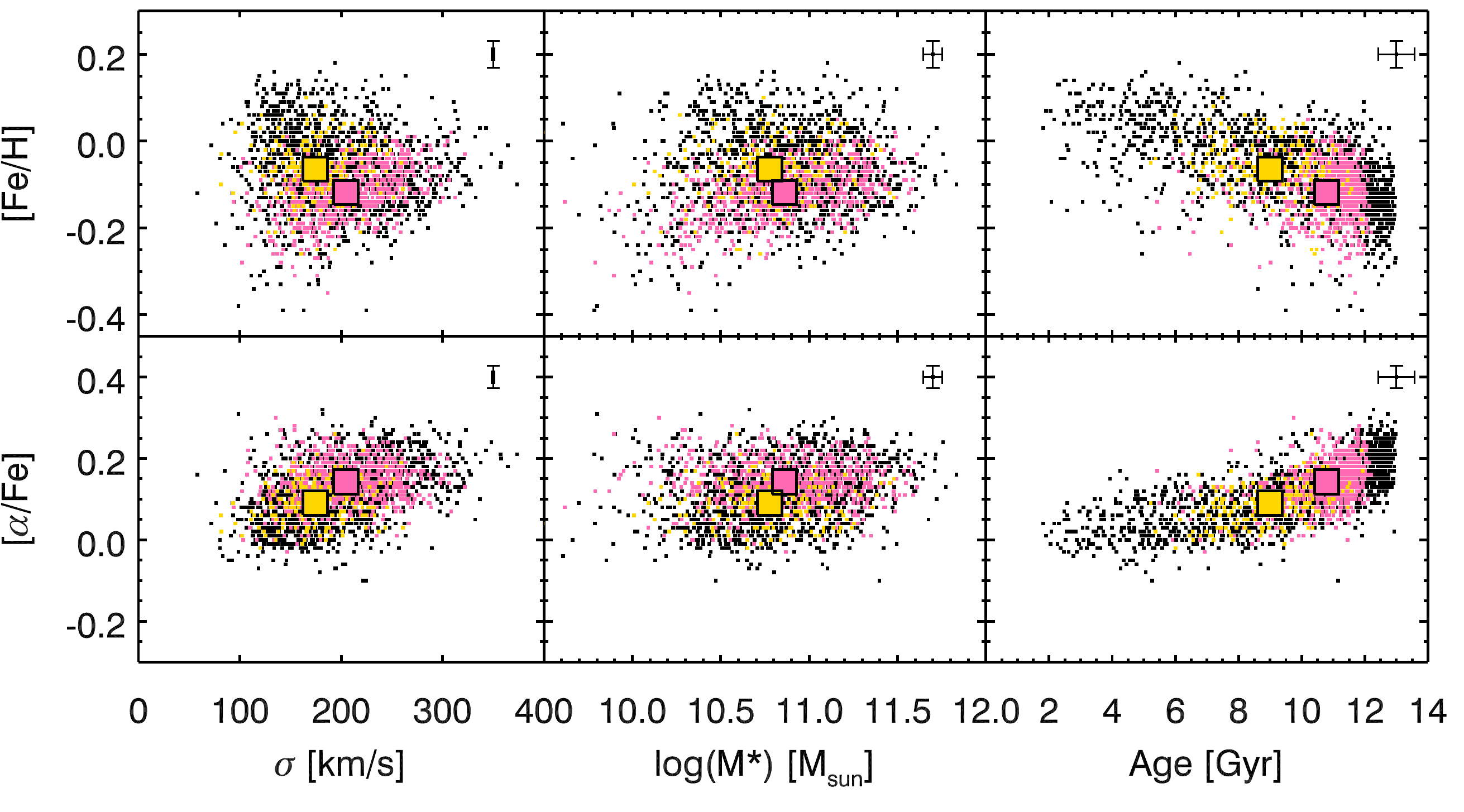}}
\end{center}
\caption[mean]{Correlations of \feh~and \afe~ with velocity dispersion, stellar mass and mean light-weighted age for our sample 
galaxies (same as Figure \ref{f:mean_quant}). Colors reproduce the classification from Figure \ref{f:distinction} into SLF-ETGs (yellow) 
and ACC-ETGs (pink). The mean for each subpopulation is also shown as a large square with black borders. Galaxies that could 
not be reliably classified are shown as black dots. }
\label{f:agemassig_dist}
\end{figure*}

Figure \ref{f:agemassig_dist} repeats Figure \ref{f:mean_quant} but adding the information about the enrichment history classification. 
It is immediately clear that on average and for our sample ACC-ETGs are significantly older than SLF-ETGs (compare also Figure 
\ref{f:sfh_MA_SE}). Also, on average SLF-ETGs have lower velocity dispersion and lower \afe~than ACC-ETGs. We come back 
to these features in the discussion Section \ref{s:disc}. 

\subsection{Environment} 

While nearly out of the focus of this paper, environment is such an important parameter for galaxy evolution that it would be 
foolish to leave it entirely out. Much work has been done on assessing the correlations between environment 
and the properties of early-type galaxies and we do not wish to simply repeat this work here, as we strongly expect to confirm 
it. Rather, in the interest of space, we concentrate only on the correlation of the old stellar population properties \feho~and 
\afeo~with environment, taking into account our classification in ACC-ETGs and SLF-ETGs. We are particularly interested to see 
whether the initial star formation event indeed depends on environment \citep[\feh~][]{renzini06}.

As a measure of environmental density we use the one by \cite{tempel12}, specifically we use the density as derived within 
8 Mpc, i.e. their parameter dens8. Their densities are based on luminosity densities around each galaxy calculated in three coordinates, 
i.e.~two projected spatial distances and one based on redshift, the latter however including advanced techniques to suppress peculiar 
motions such as suppression of the finger-of-god effect. We refer to the original paper for all the details. 

\begin{figure}[tbp]
\begin{center}
\resizebox{1\hsize}{!}{\includegraphics[]{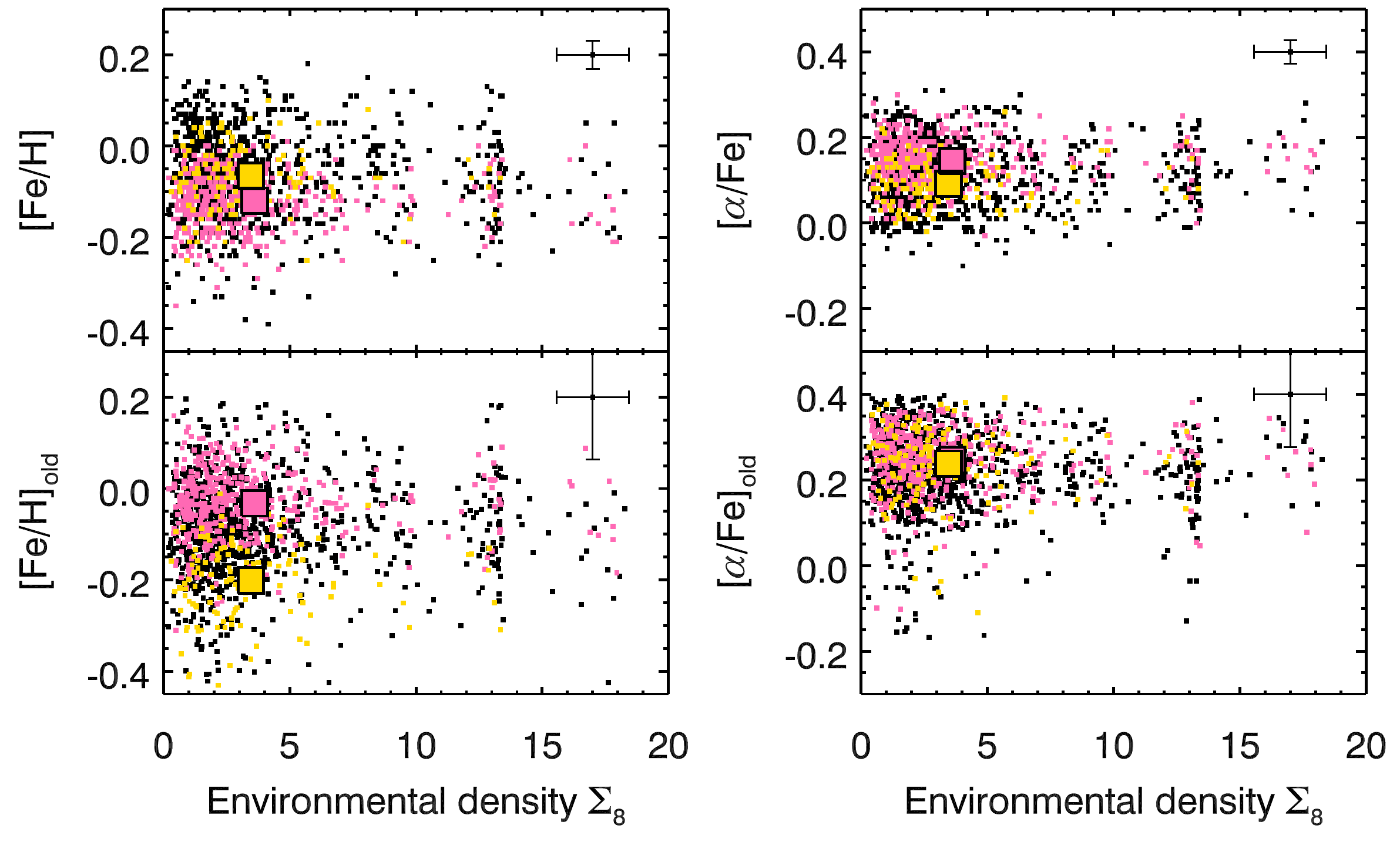}}
\end{center}
\caption[environment]{Environmental density vs.~light-weighted \feh~and \afe~(upper panels) and vs.~\feho~and \afeo~of only the 
old stars in each galaxy. The mean for each subpopulation is also shown as a large square with black borders. } 
\label{f:environment}
\end{figure}

Perhaps surprisingly we find in Figure \ref{f:environment} that the old stellar population may not be affected by environment. 
This is clearly different 
from what other authors found for the entire galaxies \citep{sanchez-blazquez06a}. It however agrees with inferences made in 
\cite{thomas10}. These authors discuss that ETG formation has occurred in two phases, where ETG galaxy formation 
depended on mass only and not on environment. \cite{rettura10} on the other hand argue that while the environment may 
modulate the timescale of the star formation history for ETGs, the formation epoch is essentially independent of environment 
and seems to depend on mass only. We here add another piece to the puzzle in saying that for the oldest stars in 
each galaxy, older than 9.5 Gyr, galaxy formation seems to have been independent of mass and environment, at least 
for the mass range probed here. 

\section{Aperture effects}
\label{s:aperture}

As discussed in Section \ref{s:intro}, the cores and the envelopes of ETGs may have assembled in different ways and 
at different times. Also, winds can redistribute elements within a galaxy \citep{pipino09}. Observationally, gradients in 
element abundances are observed for galaxies in general \citep[\feh~][]{sanchez13, perez13}. However, for ETGs the 
gradients in age and \afe~are observed to be small or null on average \citep{mehlert03, rawle08, kuntschner10}. Not so 
for the gradient in \feh, which can be significant \citep[\feh~][]{spolaor08}. The SDSS uses a single fiber to observe the 
centers of the target galaxies. The fibers fixed aperture on the sky (3\arcsec) covers a different fraction of the target 
galaxy, depending on its intrinsic size and redshift. This effect is somewhat counteracted by the effects of seeing, as seeing 
will also scatter light from larger radii into the fiber area in the focal plane. It is nevertheless clearly important to assess 
this covering fraction and the potential impact on our results. Proper aperture corrections as demonstrated in \citet{gallazzi08} 
is beyond the scope of the present paper. We also refer to the discussion in \cite{choi14} for another 
angle on the same question, albeit they work on a larger redshift range. 

\begin{figure}[tbp]
\begin{center}
\resizebox{1.0\hsize}{!}{\includegraphics[]{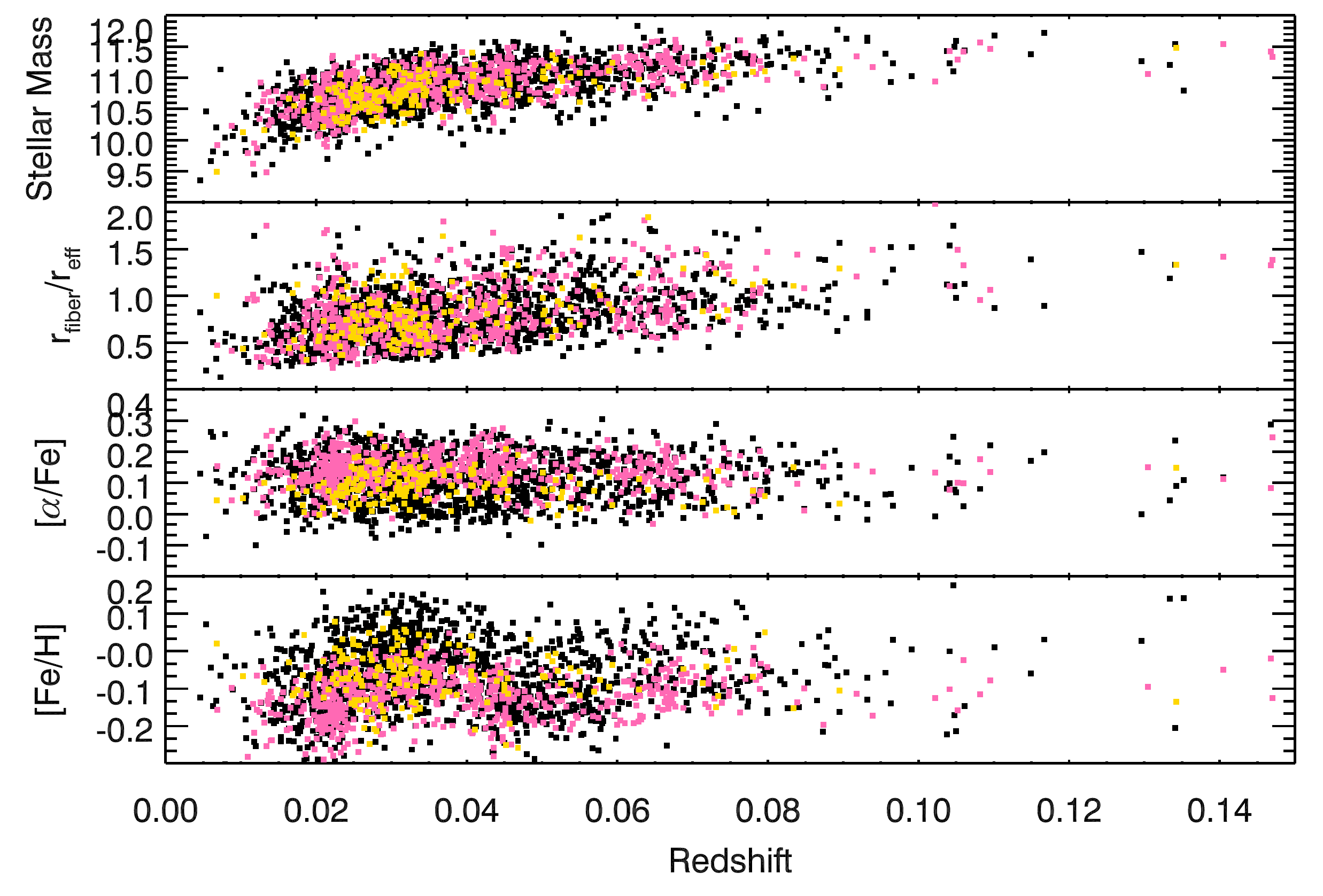}}
\end{center}
\caption[aperture]{
The dependence of key physical properties of our sample galaxies on redshift. From top to bottom: 
stellar mass, \afe, age and ratio between radius covered by fiber and effective radius of the galaxy. 
Colors reproduce the classification from Figure \ref{f:distinction} into SLF-ETGs (yellow) 
and ACC-ETGs (pink). }
\label{f:aperture}
\end{figure}

Figure \ref{f:aperture} shows the dependence of key physical properties on redshift for our sample. In the uppermost panel 
the effects of apparent magnitude selection are obvious in that we sample more massive objects at larger distances 
from Apache Point Observatory. The next panel shows the dependence of the covering fraction on redshift in terms of the ratio 
between the fiber size and the galaxies effective radius. We use the effective radius from the SDSS pipeline surface brightness 
profile fit with a de Vaucouleurs profile. It illustrates that two effects nearly cancel each other: on the one hand, the higher the 
redshift, the larger the area covered by a fiber is physically smaller on the galaxy. On the other hand, the galaxy itself 
will tend to be more massive 
and therefore larger. The covering fraction thus depends \emph{less} on redshift than the stellar mass. The next panel 
shows the expected non-dependence of \afe~on redshift, given that \afe~varies little with mass and galaxies have flat 
gradients if any. The final panel shows that \feh~depends in an interesting way on redshift. Because of the stronger 
gradients and stronger dependence of \feh~on mass it is nontrivial and beyond the scope of the present paper 
to understand this. 

\begin{figure}[tbp]
\begin{center}
\resizebox{1.0\hsize}{!}{\includegraphics[]{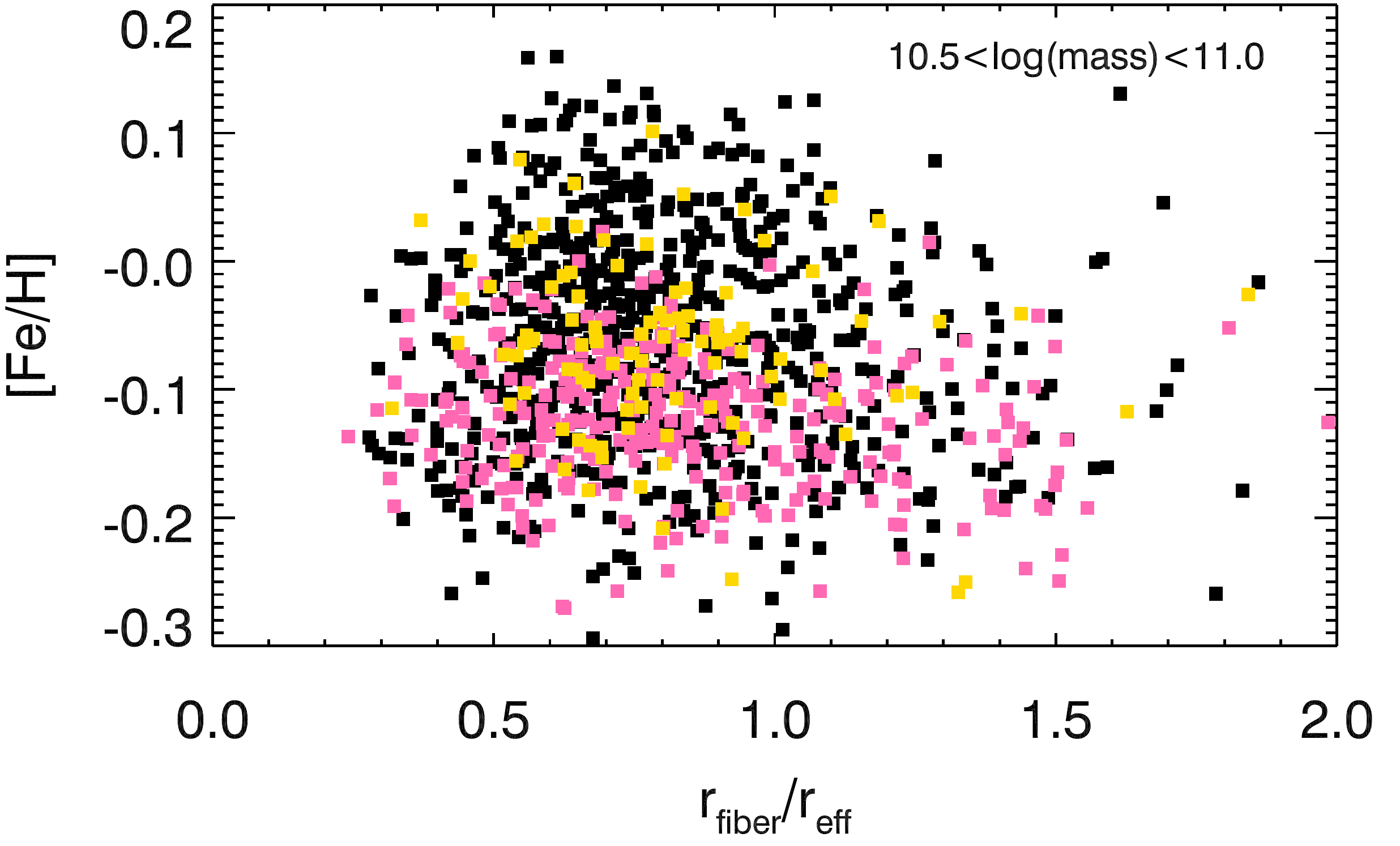}}
\end{center}
\caption[aperture_bias]{Dependence of \feh~on coverage fraction at fixed mass. No bias is obvious. 
Colors reproduce the classification from Figure \ref{f:distinction} into SLF-ETGs (yellow) 
and ACC-ETGs (pink). }
\label{f:aperture_bias}
\end{figure}

The crucial test, however, is 
shown in Figure \ref{f:aperture_bias}, where we show the dependence of \feh~on covering fraction at a fixed mass 
(here chosen to be 10.5$<$log(stellar mass/\msun)$<$11.0). There is no obvious bias to be seen, besides a potentially 
spurious tendency to have less \feh-rich objects at larger covering fraction. We conclude that tests on possible 
aperture biases are not conclusive enough to endanger any of the fairly general conclusions of this paper. 

Any further investigation of aperture effects, in particular as concerning the resolved SEHs, i.e.~ the potentially different 
biases on the old and intermediate-age stars) will have to await analysis of the currently on going integral field spectroscopy 
surveys such as CALIFA \citep{sanchez12}, SAMI \citep{allen15}, and Manga \citep{bundy15}. 

\section{Discussion}
\label{s:disc}

\subsection{Comparison of average quantities to earlier observational work}
\label{s:earlier} 

\begin{figure}[tbp]
\begin{center}
\resizebox{1\hsize}{!}{\includegraphics[]{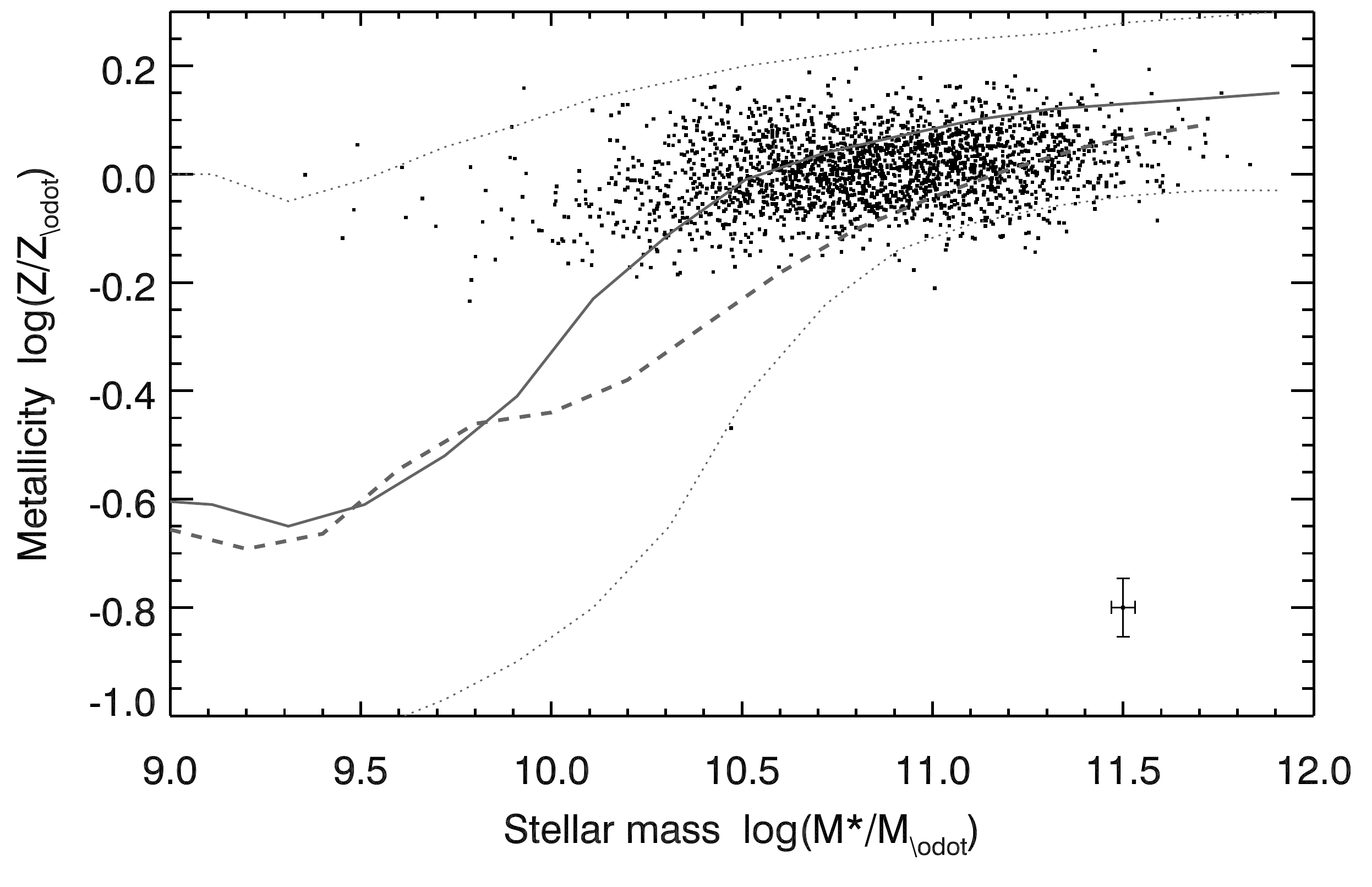}}
\end{center}
\caption[mmr]{Locus of our sample galaxies as compared to the mass-metallicity relations of \cite{gallazzi05} 
(solid line, dotted lines are 1 $\sigma$ scatter) and of \cite{gonzalezdelgado14} (dashed line). While there is general 
agreement, the clear difference in slope within the mass range of our galaxies is interesting and may point to sample 
selection effects.  } 
\label{f:mmr}
\end{figure}

A simple test is to situate our galaxies on the stellar mass -- stellar metallicity relation as defined by \cite{gallazzi05} and 
\cite{gonzalezdelgado14} as done in Figure \ref{f:mmr}. The overall match is satisfactory, given the entirely different analysis 
methods. Note though that we use the same data as for the Gallazzi relation (although our sample is smaller) and we use 
their stellar masses. Any difference between the two is therefore directly tied to the derivation of the metallicities and 
abundance ratios. Of particular interest is the difference in slope. Both the \cite{gallazzi05} and the \cite{gonzalezdelgado14} 
relation imply a significant slope at masses around $10^{11}$ \msun, where our analysis seems to imply none. This 
slope is confirmed for the case of ETGs alone in \cite{gallazzi06}. The absence of a slope in our analysis may point 
to a sample selection effect at the low mass end of our sample or may be related to the overall offset in total metallicity. 

Another interesting difference to earlier studies is the range covered by the \afe~parameter. None of our sample galaxies 
scatters above 0.3 in its light-weighted average \afe. While some of the galaxies with higher \afe~in earlier work 
\citep[\feh~][]{thomas05, kuntschner10, johansson12} may be due to observational scatter, for the stacked spectra 
of \cite{graves10}, this does surely not apply. Older models did not take the evolutionary effects of \afe~into account 
and simplified the corrections for the spectral effects. They would thus underestimate the Mg feature at given 
$\alpha$-enhancement \citep{coelho07, vazdekis11}. 
Really pinning down the cause of the offset is out of the scope of the present paper though. 

There are many works in the literature reporting correlations between \afe~and either $\sigma$ or M*. \cite{thomas05} find 
slopes of 0.28 for \afe~vs.~log($\sigma$) and 0.06 for \afe~vs.~log(M*/\msun). \cite{trager00} on the other hand find a steeper 
slope of 0.33 for \afe~vs.~log($\sigma$), similar to \cite{nelan05} who quote 0.31 $\pm$ 0.06 for the same slope. 
Finally, \cite{bernardi06} quote a slope of 0.32, while \cite{graves07} give 0.36. Our slope of 0.26 seems to lie at the low 
end of the scatter in the literature, which is probably tied to the lower maximum \afe~reached by our analysis. 

A correlation that is more rarely plotted is the one of \afe~and \feh~with age (Fig.~\ref{f:mean_quant}, right panels). Correlations 
of age and \afe~were shown in \cite{graves10} from stacked spectra (their Fig.~4) and per galaxy in \cite{kuntschner10} (their 
Fig.~6)\footnote{\cite{de-la-rosa11} have shown a correlation between the percentage of mass in old stars and \afe, but that 
is more difficult to compare to our results.}. In both cases the change of slope at an approximate age of 9 Gyr 
is visible, while the overall normalization in \afe~seems shifted higher as compared to our results by 0.05 to 0.1 dex. 
Due to the nature of the stacked spectra in \cite{graves10} and the small sample size in \cite{kuntschner10} we cannot 
comment on a comparison of the scatter around the mean relation. Although a direct comparison is difficult because of 
the different \afe~definition, a change of slope was also seen in \citet{gallazzi06}, albeit at ages of $\sim$6 Gyr, instead of 
9Gyr.We also note that these results are in contrast to the earlier work of \citet{jorgensen99} who found no correlation between \afe~and age. 

For the same correlation, it is important to heed the warnings of \cite{thomas05}, who discuss the importance of degeneracies 
when using age as a parameter. They find that for index analyses \feh~and \afe~are anti-correlated with age. As our 
dependency of \afe~on age is positively correlated with age, the observed slope cannot be due to degeneracy. For \feh~
we find that the slope of the degeneracy is approximately -0.05 dex/Gyr (see Fig.~\ref{f:degen}), whereas the slope of the 
relation is -0.02 dex/Gyr. Furthermore our relation extends over 10 Gyr, which is a much larger range in age than what 
would arise if all galaxies had the same age and would extend on an age-\feh~relation purely due to a degeneracy. We thus 
conclude that the age-metallicity degeneracy does not give rise to the observed correlation. 

These correlations of \feh~and \afe~with age explain the findings by \citet{sanchez-blazquez06a}, in which metallicity as measured by 
different indices (Fe, MgB) seemed to behave differently as a function of age. Fe-sensitive indices are a reasonable proxy of 
\feh, whereas the MgB indices depend on age, \afe~and \feh, thus likely washing out the pure \afe~signal as a function of age. 

We have shown in this paper that the age -- \afe~relation can be put on a firm foundation using the advances presented here. 
\citet{yates13} show that even the stellar mass -- \afe~relation alone is a good constraint on the delay time distribution of SNe Ia. 
We suggest that the age -- \afe~relation should provide additional interesting constraints, especially when considering the 
two regimes with different slopes. 

Finally, we caution that results such as those in \cite{trager08} seem to indicate that while many of the correlations we 
discuss are true on average over large samples, every single structure or galaxy cluster may have its own peculiar history. 

\subsection{Comparison between SLF-ETGs and the Milky Way} 
\label{s:scenario}

\begin{figure}[tbp]
\begin{center}
\resizebox{1\hsize}{!}{\includegraphics[]{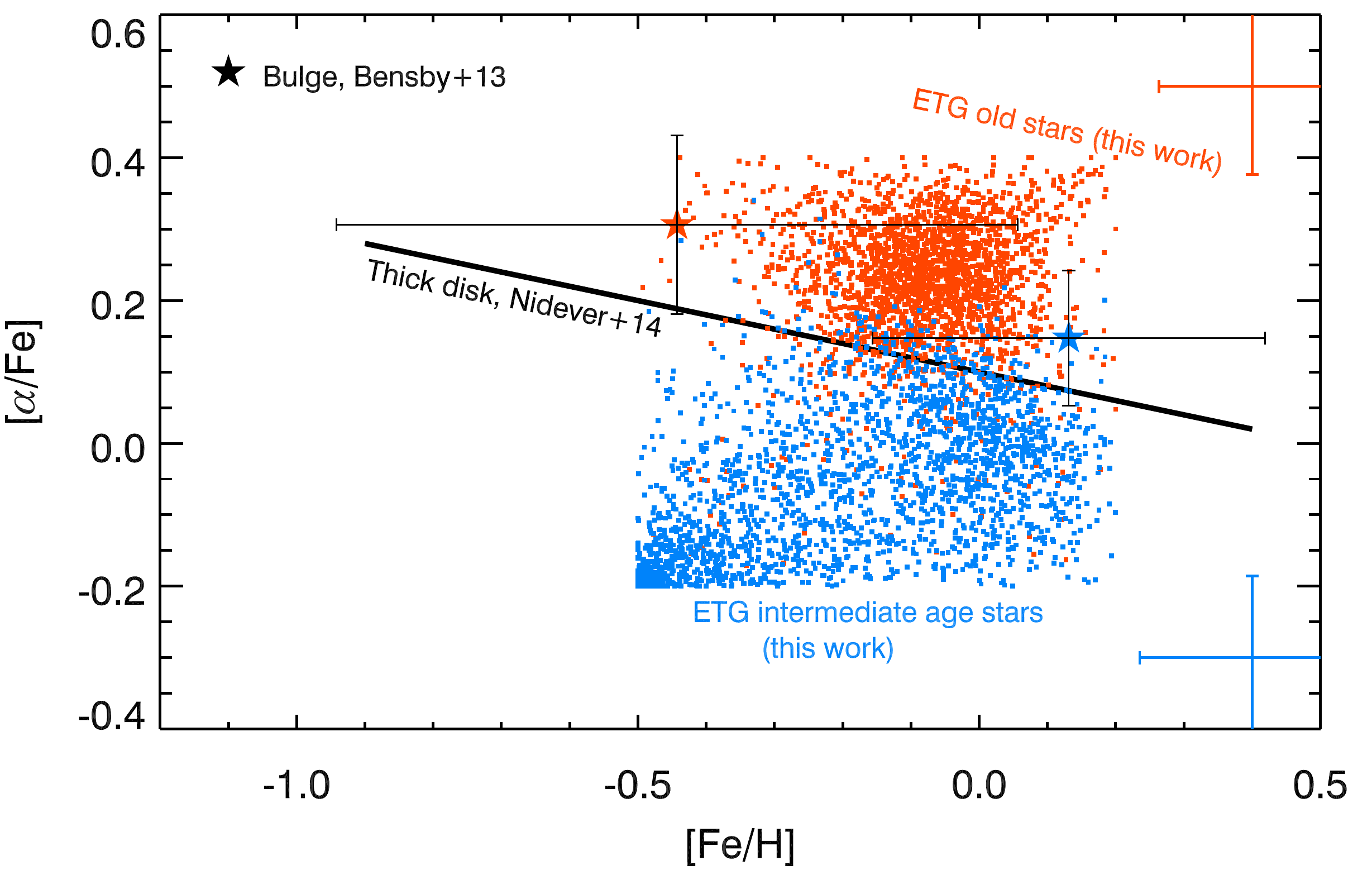}}
\end{center}
\caption[knee]{Comparison of the old and intermediate-age stellar populations in ETGs with the properties of the stars in the 
Milky Way. There is a fundamental difference in the fact that for ETGs we measure the luminosity weighted average of the properties 
over all stars in the galaxy. For the Milky Way on the other hand, the averages are over single stars and not all types of stars contribute 
to the average. } 
\label{f:knee}
\end{figure}

We have been able to diagnose for the first time the physical properties of old and intermediate stellar populations of ETGs. 
This means that for the first time we can compare the properties of these sub-populations to measurements otherwise 
only available for very nearby galaxies, mostly within the local group. We attempt this in Figure \ref{f:knee}, which reproduces 
the cartoonish expectations of Figure \ref{f:didactic}, but this time with real data. 

The two systems in the Milky Way that can be compared to ETGs in a more meaningful way are the thick disk above the plane 
(where contamination from the thin disk is minimal) and the bulge. For the thick disk we plot the fiducial line 
\afe~= $-0.2 \times $\feh~$+ 0.10$ from \cite{nidever14}. We also add the Magnesium abundances of the bulge stars older and 
younger than 9.5 Gyr from \cite{bensby13}. These represent bona-fide bulge stars and due to the reliable ages from micro-lensing 
we are able to apply the same selection as for the old stars in our ETGs. For clarity we have not plotted any values for the thin 
disk of the Milky Way, which shows continuous star formation until today. It would lie between \feh~= -1 and 0 and around 0.0 in 
\afe. 

\begin{figure}[tbp]
\begin{center}
\resizebox{1\hsize}{!}{\includegraphics[]{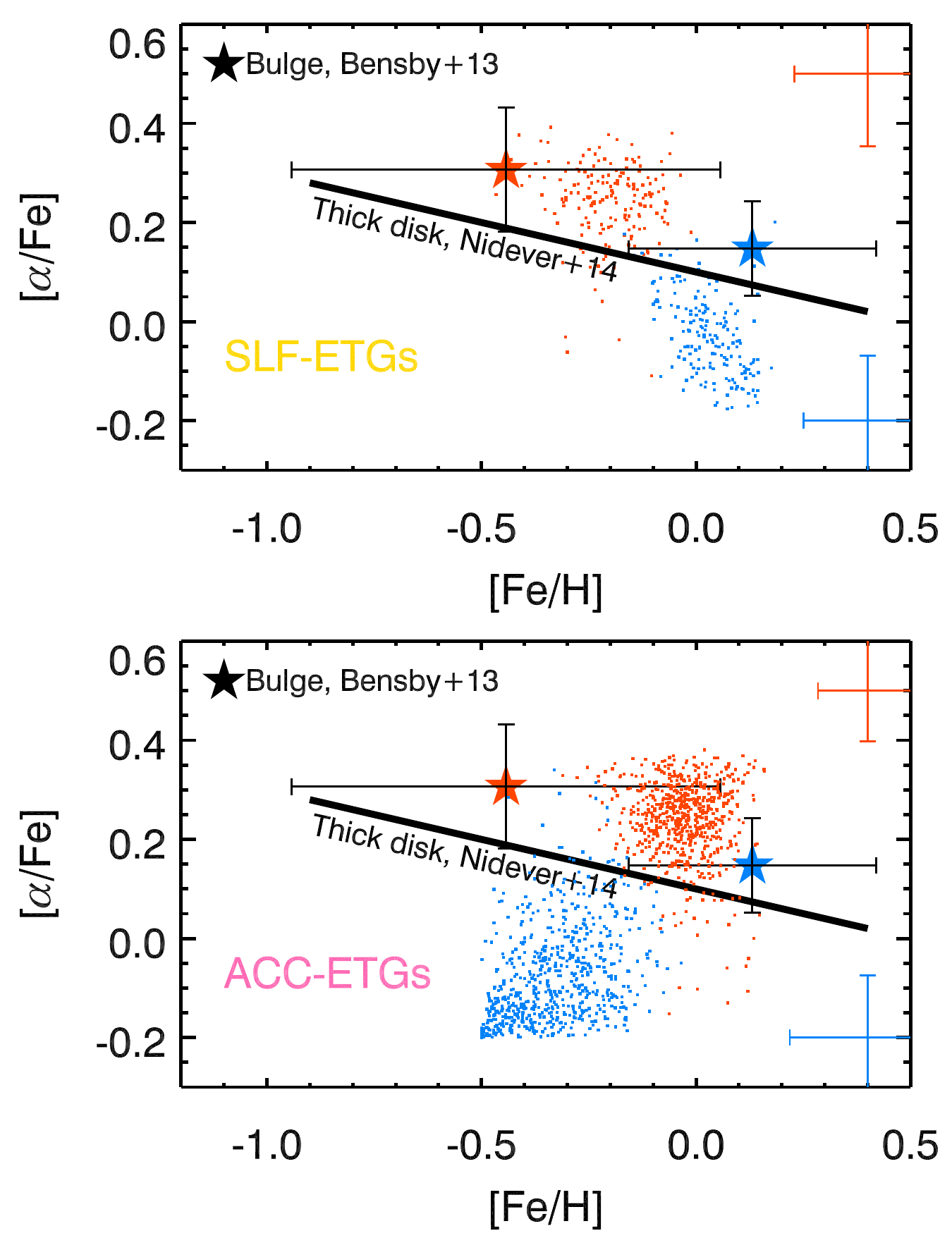}}
\end{center}
\caption[knee_SE_MA]{Old and intermediate age stellar populations in ETGs separated by the classification according 
to their enrichment history. SLF-ETGs (upper panels) are consistent with self-enrichment, while ACC-ETGs must have suffered 
either strong merging/accretion or strong internal winds. The properties of the merged galaxies can be tentatively read of 
as the properties of the intermediate age population, i.e. low in \afe~and lower in \feh~than the main body of the galaxy. } 
\label{f:knee_SE_MA}
\end{figure}

When looking at Figure \ref{f:knee} it is important to appreciate the difference in the probed stars. For the ETGs analyzed in this 
work each red dot represents the luminosity weighted average abundance values of all stars older than 9.5 Gyr in the galaxy. 
For the bulge and the thick disk, each star for which abundances have been measured contributes equally to the locus that 
is reported in this figure, i.e.~we compute a simple mean. We proceed to formulate some basic conclusions that are robust 
against possible selection effects, given the very different methods to obtain the data in MW and ETGs.

It is immediately clear from the figure that normal, massive ETGs differ wildly in their chemical enrichment history from both 
the thick disk of the Milky Way and the bulge of the Milky Way. Obviously the initial star formation event in ETGs must have 
been much more efficient in forming stars than even the stars of the thick disk or the bulge. Our statement is complementary 
to the findings of \cite{tang14} who find that the bulge is closer to ETGs than normal disk stars in its enrichment history. However, 
here we emphasize that the way in which enrichment proceeds is different, while \cite{tang14} find that the result from the 
enrichment (i.e.~the different element abundances) is different. To gain a bit more insight into the variety of manifestations of this 
fact in our sample of ETGs, we reproduce Figure \ref{f:knee} for the two classes of ETGs defined in the present paper, namely 
SLF-ETGs and ACC-ETGs. In Figure \ref{f:knee_SE_MA} we are particularly interested in the SLF-ETGs which at least are 
compatible with self-enrichment and a causally connected stellar enrichment history. Indeed, we find for these galaxies 
that the intermediate age stellar populations follow a relation between \fehi~and \afei~that is reminiscent of those followed 
by the thick disk and bulge stars, albeit with a steeper slope and extending to lower values of \afei. Very low values of 
\afe~are indeed expected for the stars formed after a long enrichment history \citep[see \feh~][their Fig.~5]{minchev13}. 

If indeed, two different processes (self-enrichment vs.~merger/accretion/winds) leave their imprint on the abundances of the 
intermediate age stars in ETGs, then it is to be expected that both processes act in all galaxies. SLF-ETGs and ACC-ETGs would 
then mostly differ in which of these two processes dominate, while the other would still play a role. This might be the 
explanation for the steeper gradient in the \afei~vs.~\fehi~relation as compared to the bulge stars. 

While such comparisons between wildly different galaxies that have been analyzed in wildly different ways should be taken 
with caution, our results are also consistent with those of \cite{silchenko93, proctor02}. Both of these author groups find that the bulges 
of disk galaxies are on average younger and less enhanced in $\alpha$-elements than ETGs. In particular, \cite{proctor02} 
state that spiral bulges show a positive correlation between velocity dispersion and metallicity, as expected from a simple 
scenario in which mass builds up slowly over a Hubble time, whereas ETGs show an anti-correlation as also found in our data. 
That our data do not support the presence of a clearly identifiable knee as we are used to from Milky Way data is 
consistent with this finding of strongly diverging processes in the chemical evolution history of late and early type galaxies. 

Finally, we note that we are unable to directly confirm the predictions by \cite{lackner12}, as in these simulations the 
accreted stars are $\sim$2.5 Gyr older, and $\sim$0.15 dex more metal poor on average than the stars formed in situ. This does not 
seem to be compatible with the distribution in \feh~vs.~\afe~space we describe here. On the other hand, \citet{navarro-gonzalez13} 
point out that it is exactly these older and metal-poor stars that are in the outskirts of galaxies, thus providing a potential way out, 
as our SDSS spectra overweight the inner parts of galaxies.

\subsection{On prolonged star formation in SLF-ETGs and the properties of minor mergers in ACC-ETGs}
\label{s:mergers}

\begin{figure}[tbp]
\begin{center}
\resizebox{1.0\hsize}{!}{\includegraphics[]{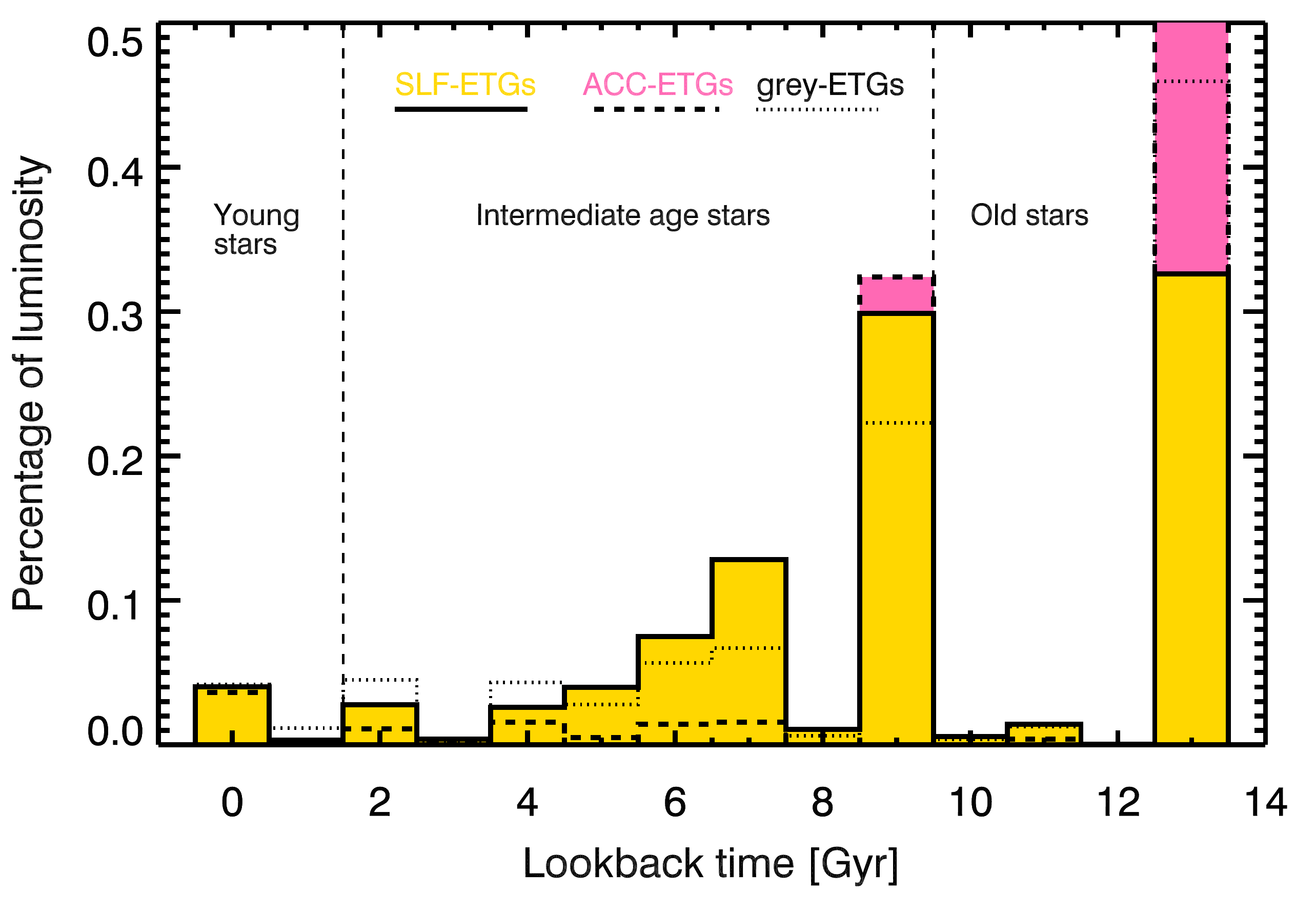}}
\end{center}
\caption[sfh_MA_SE]{Average star formation history of galaxies in the three classes as classified by their enrichment history. 
The figure shows the present day contribution of each stellar population to the total luminosity of the galaxy. All galaxies contribute 
equally to the average within each class, i.e. the total luminosity for each galaxy has been normalized to one before averaging. }
\label{f:sfh_MA_SE}
\end{figure}

Figure \ref{f:sfh_MA_SE} shows the star formation histories (i.e.~the SEHs marginalized over all abundances) for the three different 
classes of galaxies introduced in this paper. We remind the reader that there are some caveats to this classification in terms of the 
exact location of the age boundary used to separate old from intermediate age stars. The following discussion is, however, insensitive 
to these caveats. 
A general feature of Figure \ref{f:sfh_MA_SE} is that the percentage of intermediate age stars is much higher in SLF-ETGs than in 
ACC-ETGs. This would seem to indicate that mass growth through late in-situ star formation can be significant. On the other hand, 
late mergers only very rarely seem to contribute sufficient mass to significantly change the overall SFH of the ETGs. This behavior has 
been called quasi-monolithic (or early hierarchical) scenario in the literature \citep{merlin12}. For the moment this inference 
does not contradict a picture in which significant late mass growth occurs through minor mergers, as these are 
expected to be deposited at larger radii, outside the coverage of the SDSS fiber. Also, this finding is 
consistent with the recent results of \cite{hahn14} in which the authors find from entirely different lines of evidence that most 
of the ETGs stopped forming stars due to internal processes and not due to external influences. 
Our data, however, are in tension with simulations such as those by \cite{lackner12}, in which the authors found that minor mergers 
contribute between 15\% and 40\% of the total mass and that the percentage of mass brought in by minor mergers increases with 
stellar mass. We do not find that accretion increases with stellar mass, but rather that in-situ, prolonged star formation 
\emph{decreases}  with stellar mass. 

If, indeed, the intermediate age stars in ACC-ETGs mostly come from merging over a Hubble time, the properties of these 
stars should carry information on the properties of the galaxies in which they formed, i.e. on the properties of the minor 
mergers before the merging. In Figure \ref{f:knee_SE_MA} the fit results for those stars that contribute to the intermediate age 
population clearly tend towards the lower boundaries imposed by the model in terms of \feh~and \afe. In Figure \ref{f:degen} we 
showed that the degeneracy between \feh~and \afe~is more mild than for age vs.~\feh. Nevertheless, it does exist and would 
have the following effect in our case: if the real intermediate age stellar populations of ACC-ETGs were even more \feh-poor 
than allowed by our model range, the fit would compensate by lowering the \afe~further. We would thus end up exactly in 
the lowermost left corner of our model coverage -- which is what we see. Our analysis therefore tentatively indicates that 
the late mass-growth of ACC-ETGs is indeed dominated by low-mass, low-metallicity dwarf galaxies. 

\subsection{The early star formation event in SLF-ETGs vs.~ACC-ETGs}

Another interesting feature of Figure \ref{f:knee_SE_MA} are the properties of the older stars. The \afeo~and \feho~are both 
higher for ACC-ETGs than for SLF-ETGs. This implies more efficient, early star formation for ACC-ETGs. Indeed, to reach 
such high values in both \afe~and \feh~a strong starburst is needed, which may have been sufficient to drive out
 the remaining gas in the galaxies. Taking the speculation a bit further, one could imagine that ACC-ETGs would be 
 different from SLF-ETGs not in the amount of material brought in by mergers, but rather by the efficiency of their 
 early star formation event. 
 
Figure \ref{f:agemassig_dist} shows that the velocity dispersions of ACC-ETGs are on average higher than those of SLF-ETGs. 
This would imply a deeper potential well, which might have two effects: 1) increasing the efficiency of star formation, and 2) increasing 
the speed needed for wind-driven gas to leave to leave the galaxy. Whether it can be understood through simulations what causes 
effect 1) to dominate would warrant follow-up work.

\section{Conclusions}

We have analyzed a spectroscopic sample of 2286 early-type galaxies from the SDSS using state of the art stellar population models 
and fitting procedures. In particular the models we use predict the effects of age, \feh~and \afe~on the spectra of SSPs over significant 
wavelength range. This allows us to use the code {\tt paradise} to fit for the stellar enrichment history of the target galaxies without any 
prior assumption on any functional form or smoothness of this SEH. 

We conduct extensive simulations of our fitting routine and find that we can reliably recover the luminosity-weighted average properties 
of galaxies. We also find that we can reliably recover the \afe~and \feh~of the stars older and younger than 9.5 Gyr separately. We term 
those properties \feho~and \afeo~for the old stars and \fehi~and \afei~for the intermediate age stars. We introduce a classification of 
galaxies into those where \feho~$<$ \fehi~- $\Delta$ and term those SLF-ETGs because their enrichment histories are consistent with 
self-enrichment. ACC-ETGs on the other hand are those where \feho~$>$ \fehi~+ $\Delta$. For this paper we choose $\Delta = 0.15$. 

We conclude from our study that:

\begin{itemize}

\item We confirm earlier work in that the \feh~and \afe~parameters are correlated with the mass and velocity dispersion of ETGs. 
However, we find that the strongest relation is between \afe~and age, as expected from basic chemical evolution\footnote{Because 
the late onset of SNeIa enrichment is such a generic feature of chemical evolution of galaxies, this statement is not in tension with the 
assertion that ACC-ETGs make up the majority of the population.}. This relation 
falls into two regimes, one with steep slope of $0.03$ for galaxies with average ages above 9 Gyr and one with a shallower slope of 
$0.01$ for galaxies with average age below 9 Gyr. 

\item We empirically find both kinds of galaxies, i.e.~SLF-ETGs and ACC-ETGs in our sample. For our sample, bona-fide 
SLF-ETGs make up only 8\% of our sample. However, we find that for all galaxies 
independent of type \afeo$>$\afei, as expected from basic chemical evolution. SLF-ETGs are on average younger and less 
$\alpha$-enhanced than ACC-ETGs. 

\item When studying the old stars in ETGs separately, we observed very high \afe~ratios (above 0.3) and high \feh~ratios (above 
0.). On the other hand, and within our uncertainties, we find no evidence for a dependence of the properties of these old stellar 
populations on any other properties of their host galaxies, be it stellar mass, velocity dispersion or environment. 

\item ETGs on average differ strongly in their enrichment histories from any stellar system in the Milky Way, even from the bulge. 
In particular, the vast majority of them does not show a 'knee' in the plot of \feh~vs.~\afe, commonly observed in local group 
galaxies. This implies that for the vast majority of ETGs the old and the intermediate stellar populations are not bound together 
by a common enrichment history, or in other words, the stars younger than 9.5 Gyrs are likely to have been accreted through minor 
mergers. Whether the accreted material came in the form of stars or gas (that later formed stars) cannot be distinguished easily. 

\item The properties of the intermediate age stars in ACC-ETGs indicate that mass growth through late (minor) mergers is dominated 
by galaxies with low \feh~and low \afe, therefore presumably also low mass galaxies. 

\item We have physically and technically motivated the choice of boundary between old and intermediate age stars at 9.5 Gyr. However, 
the relative percentages of SLF and ACC-ETGs obtained in this paper do depend on that choice and further studies should 
be undertaken to determine the most physically and technically appropriate value. 

\end{itemize}

This contribution is the first one to attempt to resolve the stellar enrichment history of galaxies from integrated spectra. The method 
is both powerful and subject to uncertainties and caveats. We expect 
that further work in the future will allow us to confirm, extend and refine our results. This will allow us to compare the SEHs obtained for 
nearby galaxies with those of much larger samples, thus providing much needed comparison points and contributing 
to a united understanding of galaxy formation and evolution across all Hubble types. 

\begin{acknowledgements}

We thank the referee, Alexandre Vazdekis, for a very helpful report, which has helped to significantly improve the presentation 
of this paper. 
CJW acknowledges useful discussions with Davor Krajnovi\'c, Ivan Minchev, Anne Sansom, Ricardo Schiavon. 
CJW and PC acknowledge support through the Marie Curie Career Integration Grant 303912. 
SC acknowledges support from the European Research Council via an Advanced Grant under grant agreement no. 321323-NEOGAL
AG acknowledges support from from the European Union FP7/2007-2013 under grant agreement n. 267251 (AstroFIt) and EU Marie 
Curie Integration Grant "SteMaGE" Nr. PCIG12-GA-2012-326466  (Call Identifier: FP7-PEOPLE-2012 CIG).

Funding for the SDSS and SDSS-II has been provided by the Alfred P. Sloan Foundation, the Participating Institutions, the National Science Foundation, the U.S. Department of Energy, the National Aeronautics and Space Administration, the Japanese Monbukagakusho, the Max Planck Society, and the Higher Education Funding Council for England. The SDSS Web Site is http://www.sdss.org/.

The SDSS is managed by the Astrophysical Research Consortium for the Participating Institutions. The Participating Institutions are the American Museum of Natural History, Astrophysical Institute Potsdam, University of Basel, University of Cambridge, Case Western Reserve University, University of Chicago, Drexel University, Fermilab, the Institute for Advanced Study, the Japan Participation Group, Johns Hopkins University, the Joint Institute for Nuclear Astrophysics, the Kavli Institute for Particle Astrophysics and Cosmology, the Korean Scientist Group, the Chinese Academy of Sciences (LAMOST), Los Alamos National Laboratory, the Max-Planck-Institute for Astronomy (MPIA), the Max-Planck-Institute for Astrophysics (MPA), New Mexico State University, Ohio State University, University of Pittsburgh, University of Portsmouth, Princeton University, the United States Naval Observatory, and the University of Washington.

\end{acknowledgements}

\bibliographystyle{aa}
\bibliography{sdssetgs}

\begin{appendix}

\section{Fitting procedure}
\label{s:paradise}

To derive the star formation histories of composite stellar population spectra we use a full spectrum fitting software called 
{\tt paradise}. This software is based on algorithms described in \cite{rix92} and an earlier version was used previously 
in e.g.~\citet[][and references therein]{walcher06, walcher09}. Since then a number of new features have been added 
and we describe here the entire algorithm for the sake of definiteness.

\begin{itemize}
\item Read in observed spectra and the corresponding standard deviations. Rebin to logarithmic wavelength basis if original 
spectra are in linear. The set of observed spectra is considered to be in row-stacked spectrum format, with one axis being 
wavelength and the other axis allowing to fit any reasonable number of spectra sequentially.
\item Read in template spectra, rebin to same wavelength solution as object spectrum. Template spectra can be chosen 
freely, those used in this paper are described in Section \ref{s:models}. Spectra should come with a description file containing 
information such as age, mass, luminosity, \feh, \afe.
\item Specified pieces of the spectra can be masked from the fit, either in restframe (e.g. emission lines) or in observed frame 
(e.g. telluric features).
\item The continuum of the object and the templates can either be left untouched, or taken out. In the second case the continuum 
can be determined through a set of Legendre polynomials or through a running mean with to-be-specified width. The continuum 
can be subtracted or the spectra can be divided by it. For continuum computation reasons, a separate restframe mask can 
be supplied.
\item Use an initial template to obtain a reasonable value for the velocity and velocity dispersion of the object through 
a $\chi^2$-biased random walk. The line-of-sight velocity distribution is assumed to be Gaussian. This losvd is applied 
to all template spectra. 
\item Apply a non-negative least squares algorithm \citep[nnls,][]{lawson74} to perform the inversion onto the stellar population model 
\citep[see][for details on this step]{walcher06}.
\item Finally, repeat $\chi^2$-biased random walk to yield a best estimate for the kinematic parameters, 
starting from the best fit values obtained before. 
\item The nnls routine has the feature of preferring some templates over others, leading to spurious non-continuous star formation 
histories. Therefore one can choose to repeat the stellar population inversion many times, where the object spectrum is noised 
assuming Gaussian noise and the variance vector, the order of the templates is reshuffled and a certain percentage of the 
template set spectra is not considered anymore. This is what we call the stellar population bootstrap. The stellar population bootstrap 
allows also to derive useful rms uncertainties on any of the stellar population parameters output by the code. In practice we find that 
after 100 bootstrap runs and using 80\% of the initial template set we at the same time obtain smooth SFHs and useful errorbars. 
We believe this approach to be a straightforward alternative to regularization \citep{ocvirk06, cappellari04}. 
\item It is often unclear to what extent stellar population uncertainties can bias the velocity dispersion measurements. To address 
this, the code has the capability to repeat the kinematic fit after each step of the stellar population bootstrap. We call this the 
kinematics bootstrap. This feature is not used in the current paper. 
\item Finally the code puts out the fiducial best fit spectrum, the continuum it used to normalize the object spectrum, the associated errorbar, 
and residual. It also provides the user with two tables, one with the fiducial values of the fit and the other with the results of the bootstrap. 
Finally, the variance vector on the best fit derived from the bootstrap procedure is also provided. In this paper we generally use the 
stellar population parameters obtained from the bootstrap as these come with errorbars. Within the errorbars, these are consistent 
with the fiducial values. 
\end{itemize}

For a review of other available spectral fitting codes, the reader is referred to \citet{walcher11} and the website http://www.sedfitting.org. 

A general note on averaging may also be useful for the reader: when quoting the mean \feh~and \afe~abundances, it is 
not trivial to determine the most appropriate way to do the averaging. One could either 1) average on the "square bracket 
quantities", or 2) average on actual element abundance ratios, or 3) average on actual masses of Fe atoms M(Fe) and of 
Hydrogen atoms M(H) (alpha elements and Fe respectively for the case of \afe). We have decided to stick with possibility 
one for an observational reason: interpolating linearly two stars with \feh=-1 and \feh=0 exactly reproduces a star with \feh=-0.5 
(tested using stellar spectra from \cite{coelho05}). Because we can thus observationally not distinguish between these 
last two situations, we shall use an averaging procedure in which they both correspond to the same result. In this way 
we avoid inconsistencies and degeneracies. When attempting to compare our results to simulations, one will 
have to do the same to their derived quantities. 

\end{appendix}

\end{document}